# A lateral nanoflow assay reveals surprising nanoplastic fluorescence

Kuo-Tang Liao,[∥] Andrew C. Madison,[∥] Adam L. Pintar, B. Robert Ilic, Craig R. Copeland, and Samuel M. Stavis*

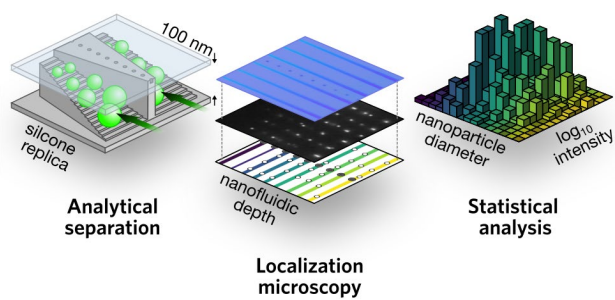

**ABSTRACT:** Plastic nanoparticles present technological opportunities and environmental concerns, but measurement challenges impede product development and hazard assessment. To meet these challenges, we advance a lateral nanoflow assay that integrates complex nanofluidic replicas, optical localization microscopy, and novel statistical analyses. We apply our sample-in-answer-out system to measure polystyrene nanoparticles that sorb and carry hydrophobic fluorophores. An elegant scaling of surface forces automates advection and dominates diffusion to drive the analytical separation of colloidal nanoparticles by their steric diameters. Reference nanoparticles, with a mean of 99 nm and a standard deviation of 8.4 nm, probe the unknown limits of silicone replicas to function as separation matrices. Innovative calibrations correct aberration effects from the microscope and the device, improving the accuracy of reducing single micrographs to joint histograms of steric diameter and fluorescence intensity. A dimensional model approaches the information limit of the system to discriminate size exclusion from surface adsorption, yielding errors of the mean ranging from 0.2 nm to 2.3 nm and errors of the standard deviation ranging from 2.2 nm to 4.2 nm. A hierarchical model accounts for metrological, optical, and dimensional variability to reveal a fundamental structure–property relationship. Intensity scales with diameter to the power of 3.6 ± 0.5 at 95 % coverage, confounding basic concepts of surface adsorption or volume absorption. Distributions of fluorescivity—an intrinsic property that we define as the product of the number density, absorption cross section, and quantum yield of an ensemble of fluorophores—are ultrabroad and asymmetric, limiting any inference from fluorescence intensity. This surprising characterization of common nanoplastics resets expectations for optimizing products, applying standards, and understanding byproducts.

**KEYWORDS:** *nanofluidics, nanoplastics, nanoparticles, fluorescence*

Colloidal nanoparticles of polymeric materials and organic chemicals are at the vanguard of commercial nanotechnology, with diverse applications ranging from instrument calibration to therapeutic delivery.[1-4] At the same time, environmental pollution from both nanoparticle products and nanoplastic byproducts is of grave concern,[5] with potential hazards of chemical sorption and tissue penetration.[6-8] Structure–property relationships of nanoplastic particles are fundamental to control in production, to depend on in application, and to understand in pollution. However, quantitation and correlation of structure and property distributions is challenging. Accurate and precise measurements of dimensional and chemical properties, often by optical proxies,[9-11] can be inefficient or impractical using conventional methods. For example, the combination of multiple methods of microscopy[12-14] encounters the limits of preparing colloidal samples and sampling many single nanoparticles, while incurring the issue of calibrating and correlating disparate sources of microscopy data. In contrast, rapid and economical measurements of nanoparticle ensembles in colloidal suspensions, such as by dynamic light scattering or fluorescence correlation spectroscopy, yield little information about structure and property distributions. Combining chromatographic methods with ensemble measurements provides access to more distributional information[15] but still obscures population heterogeneity from variability among single nanoparticles and introduces challenges of surface interactions.[16, 17]
In this way, conventional methods of nanoparticle characterization tend to achieve either quality or efficiency. For this reason, emerging methods seek to obtain both metrics of performance, often by combining optical microscopy, nanoparticle tracking, and fluidic devices.[18-23] Each method exhibits certain capabilities and limitations[24, 25] and, despite the potential of these emerging methods, measurements of structure and property distributions that accurately capture both the sample mean and root variance or standard deviation remain challenging. In the present study, we develop a nanofluidic and microscopic analogue of a lateral flow assay, achieving a new combination of quality and efficiency in colloidal metrology. We apply this assay to reveal that a ubiquitous nanoparticle sample – which is simultaneously a commercial nanotechnology, model nanoplastic,[26] and unofficial standard[27] – is hiding surprising fluorescence in plain sight. This discovery has important implications in the many contexts that involve such samples and for which an unpredictable or unreliable relationship between structural and optical properties could be problematic.

In previous studies, we fabricated the first nanofluidic devices with complex three-dimensional structures in the form of confining staircases,[28] separated polystyrene nanoparticles sorbing chemical fluorophores at staircase step-edges by nanofluidic size-exclusion from shallower steps,[29] and referred nanoparticle positions to step depths to measure diameter distributions with subnanometer accuracy and mean fluorescence intensity per step.[30] In this way, our devices function as analytical separation matrices and nanoscale dimensional standards for use with an optical microscope. However, three critical issues limit the practical application of previous implementations of our method. First, nanofluidic devices in hard materials require a large input of fabrication and characterization but have a short lifetime in our application, which involves forcing nanoparticles into confinement at the steric limit. This limitation motivates the efficient replication of disposable devices in a laboratory context, but polymeric nanostructures, even in a production context, can be mechanically and chemically unstable.[31, 32] In particular, the elastic deformation of silicone can limit control of the structure and function of nanofluidic replicas,[33, 34] presenting an open question as to the feasibility of the transfer of our methodology from hard to soft materials. The second issue is that a high throughput of single nanoparticles across a wide field is necessary to efficiently sample a heterogeneous population. However, inaccurate placement of features by focused-ion-beam[35] machining and inaccurate measurements of position and intensity by optical microscopy[36] are common issues in both widefield patterning and imaging. Systematic effects from aberrations in each system impede the lateral scale-out of the assay and require integration of new standards and calibrations into the device and analysis.[36, 37] The third issue is that nanoparticles can adsorb nonspecifically to device surfaces,[28-30] resulting in spurious data that obscure the steric interaction of size exclusion. Surface interactions are generally challenging for manipulating nanoparticles in fluidic and chromatographic systems[38] and could be limiting for nanoplastics with unknown surface properties. These interactions are unavoidable to some extent, motivating new analytical approaches to reduce imperfect data into reliable measurements of nanoparticle size.

We address these critical issues, establishing the technological and analytical scalability of our lateral nanoflow assay in a fundamental study of nanoplastic fluorescence heterogeneity. From sample-in to answer-out, the integration of device, microscope, and analysis into a practical and reliable system is a theme of our study. We begin by exploring the replication[39-41], application, and characterization of complex nanofluidics near the atomic scale, studying the transport of a colloidal suspension in silicone staircases. Capillarity drives the steric interaction of nanoparticles with the replica structure, automating separation and enabling readout of the steric diameters and fluorescence intensities of single nanoparticles by localization microscopy. In turn, the diameter distribution of reference nanoparticles probes the functionality and validates the stability of our silicone replicas. Innovative calibrations and statistical models improve accuracy across an ultrawide field, accounting for aberrations of both the device and the microscope to reliably reduce single optical micrographs into joint histograms of steric diameter and fluorescence intensity. This approach enables comprehensive analyses that robustly discriminate size exclusion from surface adsorption, making use of varying amounts of prior information about diameter bounds. Building on the resulting accuracy and throughput, a hierarchical statistical analysis of steric diameter and fluorescence intensity reveals a fundamental structure–property relationship of a common, if unofficial, standard. Fluorescence intensity exhibits a super-volumetric scaling with steric diameter, confounding basic concepts of the proportionality of chemical sorption to surface area or volume. Distributions of fluorescivity—an intrinsic optical property that we isolate and define as the product of the number density, absorption cross section, and quantum yield of an ensemble of fluorophores interacting within the dielectric volume of a nanoparticle—are ultrabroad and asymmetric, underscoring severe limits of ensemble analysis, elucidating the unreliable fluorescence intensity of single nanoparticles, and undercutting inferences of dimensional or chemical properties from fluorescence intensity. These surprising results reset expectations for optimizing products, applying standards, and understanding byproducts involving nanoplastic fluorescence.

## RESULTS AND DISCUSSION

**Complex nanofluidic replicas.** Control of vertical dimensions is essential to our method, as step depths limit separation resolution. Therefore, we begin by testing the fidelity of our replication process (Figure 1a-d).[42] We machine complex molds in silica using a focused ion beam (Table S1) and measure surface topography by atomic force microscopy.[30] In an initial test of pattern transfer, staircase structures with subnanometer steps form inverse replicas in silicone with a fidelity of approximately 0.1 nm (Figure S1).[40] Deposition of a fluorosilane release agent increases surface roughness by up to 1 nm, elucidating a practical aspect of our nanomanufacturing process (Figure S2). Our device design is a parallel array of staircase structures that decrease in both depth and width to maintain an aspect ratio that prevents channel collapse.[33, 43] Device patterns (Table S2, Figure S3) transfer first from a silica mold (Figure 1a, e) to an inverse replica in silicone (Figure 1b) and then to a silicone replica (Figure 1c, f). In test structures that quantitate this process, depth increments of 1.8 nm ± 0.5 nm decrease on average by 0.3 nm ± 0.2 nm, whereas surface roughness increases from 0.65 nm ± 0.07 nm to 0.74 nm ± 0.07 nm (Table S3). We report uncertainties as 95 % coverage intervals. In a last test of pattern transfer, we form complex surfaces in nanoscale films of silicone (Figure S4). These results are necessary to test the subnanometer fidelity of pattern transfer and to enable future devices in thin films, but are still insufficient for our nanofluidic application.

Accordingly, we perform an operational test of nanofluidic size-exclusion using nanoparticle standards with a reference diameter histogram from transmission electron microscopy (Table 2), which we model with a Johnson $S_U$ distribution to guide the eye (Figure S5, Table S4). This test probes the unknown limit of replica structure and function through steric interaction with reference nanoparticles. Silica coverslips enclose silicone replicas (Figure 1c-d), forming nanofluidic devices, which automate the transport and separation of colloidal nanoparticles (Figure 1g-i, Scheme S1). Brief exposure of the coverslips and replicas to oxygen plasma[44] hydroxylates their surfaces to promote bonding and to increase hydrophilicity, induce capillarity of the colloidal suspension, and repulse nanoparticles with anionic surfaces, which are common in colloidal interfaces.

**Nanofluidic size exclusion.** An aqueous suspension of spherical polystyrene nanoparticles carrying boron-dipyrromethene fluorophores and having surfaces terminating in carboxylic acids, fills the silicone replicas. This sample-in process is effortless from the practitioner perspective and results in a novel transport effect, which exploits surface forces to selectively transport the suspension. While the device surfaces remain hydrophilic and are wetting, a capillary pressure drives fluid flow. As the staircase structures decrease in depth and width, the flow speed of the dispersion medium and resulting advection of nanoparticles increase as hydrodynamic interactions progressively hinder the diffusion of nanoparticles.[45] The effect is increasing Brenner numbers[46] with upper bounds of $10^1$ to $10^2$ during size exclusion, despite the increasing diffusivity of nanoparticles of decreasing size (Figure 1g-h, Table S5, Figure S6). This unusual scaling suppresses Brownian noise, elegantly solving a problem of many schemes of analytical separation. In a further test of system hydrodynamics, replica stability, and device design, we study size exclusion in channel arrays with variable depths. Nanoparticle sampling increases with inverse hydraulic resistance (Figure S7), informing the design of devices for measurements of particles with complex size distributions.

### Table 1. Nanoparticle diameters

| measurement method | experiment | relevant bounds | mean | error | standard deviation | error |
|---|---|---|---|---|---|---|
| transmission electron microscopy | – | lower and upper bounds | 99.3 nm ± 0.5 nm | — | 8.4 nm ± 0.4 nm | — |
| lateral nanoflow assay | 3 | upper bound | 99.3 nm ± 1.9 nm | 0.0 nm | 11.9 nm ± 1.7 nm | 3.4 nm |
| lateral nanoflow assay | 3 | lower and upper bounds | 100.5 nm ± 1.7 nm | 1.2 nm | 10.4 nm ± 1.2 nm | 2.0 nm |
| lateral nanoflow assay | all | upper bound | 97.0 nm ± 1.6 nm | 2.3 nm | 12.6 nm ± 0.8 nm | 4.2 nm |
| lateral nanoflow assay | all | lower and upper bounds | 98.9 nm ± 1.3 nm | 0.4 nm | 10.6 nm ± 0.8 nm | 2.2 nm |

The lower and upper bounds are 78.1 nm and 134.3 nm, respectively, from the 99.9 % coverage interval of the reference diameter histogram (Figure S1, Table SError! Bookmark not defined.).
The mean and standard deviation correspond to the first and second moments of the particle size distribution. The uncertainties of these values are 95% coverage intervals.

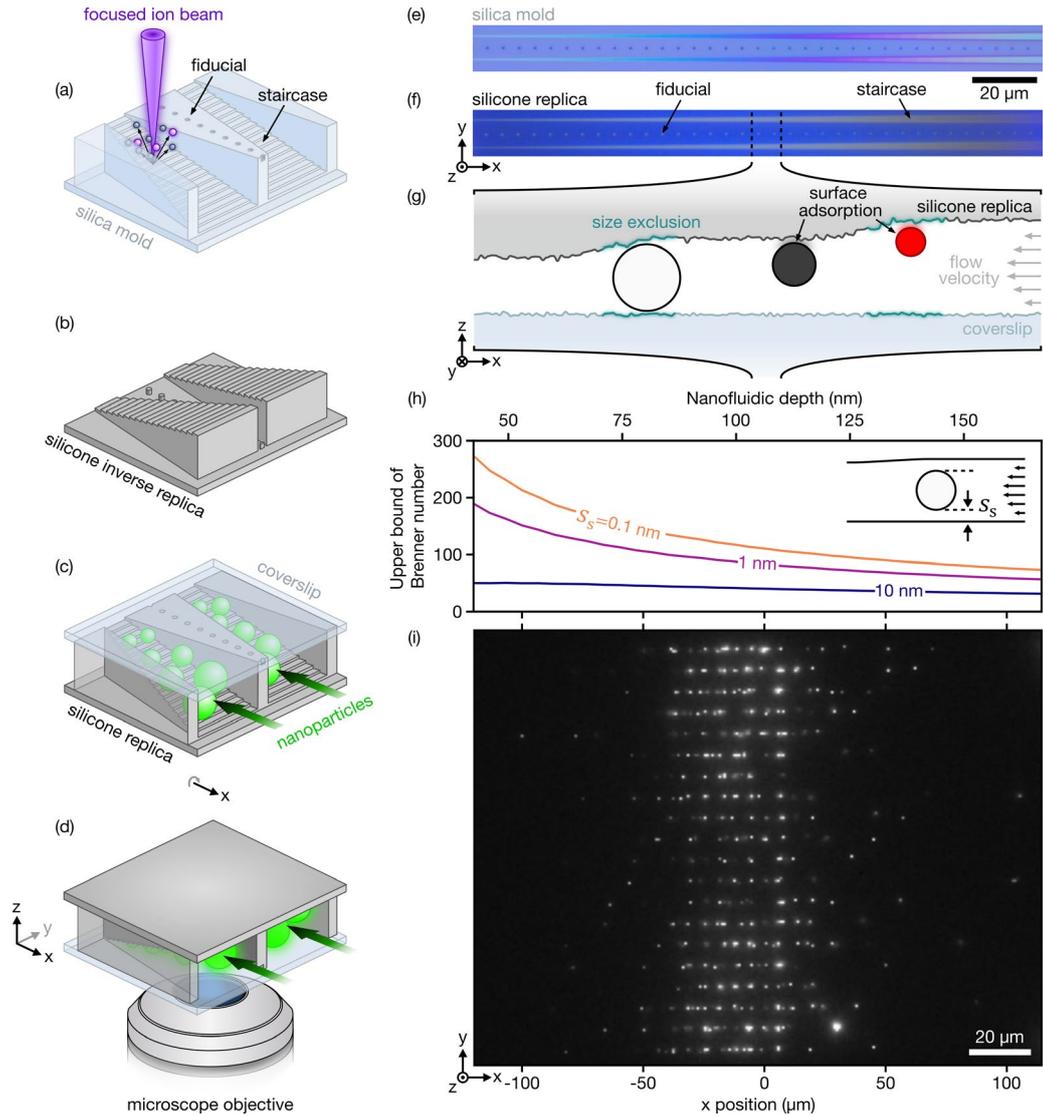

**Figure 1.** Nanofluidic size exclusion. (a-d) Schematics showing (a) a silica mold, (b) an inverse silicone replica, (c) a silicone replica with a silica coverslip, and (d) the orientation of the device with respect to the microscope objective. (e-f) Brightfield optical micrographs showing (e) a silica mold and (f) a silicone replica. (g) Schematic showing a side view of nanoparticles interacting with a nanofluidic staircase, undergoing (white) advection and size exclusion and surface adsorption either (black) outside of or (red) inside of size-exclusion regions. (h) Plot showing upper bounds of theoretical Brenner numbers for (inset) three separation distances between a nanoparticle and confining surfaces, $s_s$. (i) Fluorescence optical micrograph showing the size separation of polystyrene nanoparticles in a nanofluidic staircase.

**Measurement system calibrations.** After size separation, brightfield microscopy forms images of fiducials (Figure 1f, Figure S8), and fluorescence microscopy forms images of nanoparticles (Figure 1i). These micrographs are a rich source of data but are also rife with sources of error, including illumination nonuniformity, image proximity, widefield aberrations, and thin-film interference. We develop four calibrations to identify, correct, and study the resulting systematic effects, which are highly complex and generally dependent on position (Figure 2). This is the way to make an accurate measurement,[47, 48] integrating the device and microscope into a measurement system that achieves a high level of metrological reliability. We first summarize the four calibrations and then revisit their effects in a functional evaluation, testing sizing accuracy by a comparison of our results to the reference diameter histogram from transmission electron microscopy.

The first calibration is a flatfield correction. A featureless and autofluorescent film, with a nanoscale thickness just exceeding the vertical dimension of the deepest features of the staircase structures, serves as a reference object (Figure 2b). Imaging this object samples the focal volume under widefield illumination in a way that is comparable to that of imaging the nanofluidic device by brightfield and fluorescence microscopy. The resulting images manifest the corresponding illumination nonuniformities. Analysis of the signal intensities yields accurate flatfield corrections, which differ significantly due to the differences of illumination optics (Figure S9) and can significantly affect the results of localization and intensity analyses.[36, 49] Thicker films and surfaces with regions that are out of focus degrade the accuracy of the flatfield corrections (Figure S10), demonstrating the importance of matching the interaction of sample and microscope between calibration and experiment for accurate measurements of signal intensity in widefield imaging.

The second calibration is a point-spread-function filter. Device fiducials and sparse nanoparticles from control experiments provide reference images for both microscopy modes. Symmetric Gaussian models[50] approximate the images of fiducials that begin to resolve and saturate in brightfield micrographs (Figure S11, Table S6)[51] and subresolution nanoparticles in fluorescence micrographs (Figure S12).[36] Beyond demonstrating the counterintuitive concept of localization above the saturation limit of an imaging sensor with subnanometer accuracy, Gaussian fitting enables filtration of data to preclude analysis of the images of multiple nanoparticles that are in proximity near the resolution limit, which is an advantage over direct summation of signal photons. Analysis of the standard deviations of Gaussian approximations of sparse nanoparticle images, which resemble the point spread function with varying defocus, yields a threshold value of 175 nm at 95 % confidence to discriminate between the images of sparse and proximate nanoparticles (Figure 2d-e, Figure S13). This maximizes sampling and minimizes errors from image overlap. We confirm that integration of the Gaussian model fits underestimates signal intensity[52, 53] and we establish a reliable proportionality between Gaussian integration and direct summation of signal photons (Figure S14), yielding robust measurements of fluorescence intensity. This analysis shows a new benefit and reliability of the default localization analysis involving a symmetric Gaussian approximation of the point spread function.

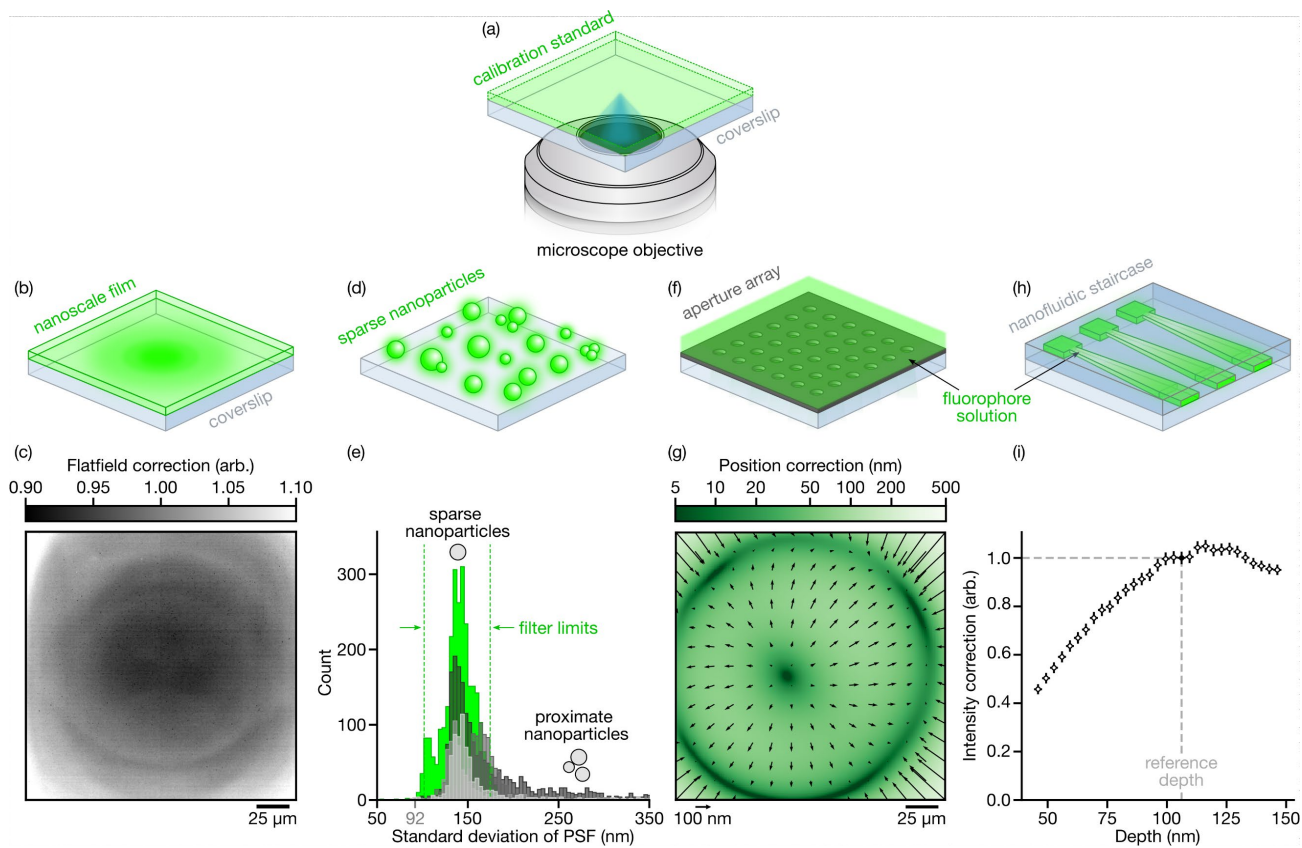

**Figure 2. Measurement system calibrations.** (a) Schematic showing a generic standard for microscope calibration. Specific implementations follow. (b) Schematic showing a fluorescent film with a thickness just exceeding the device depth to form a focal volume for flatfield correction. (c) Plot showing the flatfield correction for epi-illumination fluorescence micrographs. (d) Schematic showing a control sample of sparse nanoparticles to quantify limits of the standard deviation of the point spread function (PSF). (e) Histogram showing standard deviations from (green) control measurements and (grayscales) experimental results in nanofluidic devices. (Dash green lines) A 95 % coverage interval of the standard deviation of sparse nanoparticles, ranging from 102 nm to 175 nm, filters emitter images with larger standard deviations due to proximate nanoparticles in nanofluidic devices. (Gray tick mark) The theoretical value of the standard deviation is approximately 92 nm[55] (Figure S11). Experimental values exceed 92 nm due to field curvature and defocus. (f) Schematic of an aperture array and fluorophore solution to provide reference positions for fluorescence micrographs. (g) Vector plot on a linear scale and color map on a logarithmic scale showing the position corrections of nanoparticles. The field dependence results from distortion among other optical aberrations. (h) Schematic of a nanofluidic staircase array and fluorphore solution to quantify interference effects in dielectric films. (i) Plot showing intensity correction for fluorescence intensity as a function of nanofluidic depth. Normalization is with respect to the intensity at the reference depth (black circle) of 106 nm ± 3 nm. Horizontal and vertical bars are increments of nanofluidic depth and 95 % coverage intervals of the intensity correction.

The third calibration is a position correction. An aperture array that we fabricated previously by electron-beam lithography provides reference positions to correct errors due to microscopy distortion, among other aberration effects of the two imaging modes, improving localization accuracy by orders of magnitude across our imaging field.[36] Analysis of apparent positions of apertures at the different wavelengths yields position corrections that differ significantly for fiducials and nanoparticles (Figure 2f-g, Figure S15). Depending on position in the imaging field, the correction magnitude ranges from 2 nm to 130 nm and has the net effect of increasing the number of nanoparticles in regions of size exclusion by approximately 20 % (Table S7). After correction of apparent positions, analysis of fiducial positions reveals actual fabrication distortion of the device pattern at a similar scale (Figure S15e-g), which we account for in our statistical measurement model. In comparison, nanoparticle images have smaller values of theoretical localization precision, decreasing from 10 nm to 1 nm as fluorescence intensity increases (Figure S15h), showing the potential for false precision and overconfidence in apparent positions. We revisit the effects of the position correction, which is essential for areal scale-out of the assay.

The fourth calibration is an intensity correction. A fluorophore solution, with an emission spectrum closely resembling that of our nanoparticles, fills the nanofluidic staircases to reveal interference effects in complex films of dielectric materials.[30] The resulting relationship between nanofluidic depth and fluorescence intensity allows linearization of fluorescence intensity in replicas through an intensity correction of appreciable effect (Figure 2h-i, Figure S16). Replica autofluorescence is uniformly low,[54] facilitating this calibration and enabling future detection of faint signals from colloidal nanoplastics and molecular adsorbates.

**Statistical measurement model.** With all four calibrations complete, it is possible to accurately refer nanoparticle positions to the nanofluidic depths of the device to measure apparent diameters (Figure 3a-c). Several dimensions and their uncertainties are critical to this registration. First, a mean step depth of 3.3 nm ± 0.4 nm sets the separation resolution and bin width of diameter histograms (Figure 3b-c). This value is less than 40 % of the standard deviation of the reference diameter histogram, ensuring sufficient resolution for sampling the diameter distribution. Depth uncertainties of the device, from surface roughness of the replica and coverslip and various calibrations of our atomic force microscope, yield diameter uncertainties for single nanoparticles that have 95 % coverage intervals ranging from 2.3 nm to 2.5 nm. Last, the widths of step edges, where steric interaction between nanoparticles and replicas occurs, have a mean value of approximately 560 nm.

In a new model of size exclusion, Monte-Carlo simulations account for a comprehensive set of dimensional parameters and their uncertainties (Figure 3a-b, Table S8, Figure S17). A quadratic polynomial models each of the 36 columns of three fiducial positions, which mark step edges in the devices. This parabolic approximation of fabrication distortion allows for curving size-exclusion regions (Figure 3b), without fitting errors but with uncertainty propagation from step-edge width, nanofluidic depth, and fiducial localization, in order of decreasing magnitude and among other effects (Table S8). This new analysis is conceptually different from the assumption of device linearity in previous studies[28-30] and is essential to areal scale-out of the assay. Integration of more fiducials and extension of the polynomial models to higher order enables analysis of arbitrary device curvature, whether non-linear by design or by distortion, across a wide field. Ultimately, our Monte-Carlo simulations yield empirical distributions of the positions and dimensions of size-exclusion regions and positions of nanoparticles, which reduce to 95 % coverage intervals to reject outlying nanoparticles (Figure 3b). Subsequent combination of position and diameter data allows rejection of pairs of nanoparticles with apparent separations less than the sum of the apparent radii in each size-exclusion region (Figure S18, Table S9).

This rigorous statistical modeling of the separation and measurement indicates that approximately 30 % of nanoparticles are in size-exclusion regions and 70 % are in interstitial regions during the measurement (Table S10). This analytical yield begins to quantify the hindrance by surface adsorption of nanoparticle advection toward size-exclusion regions. The results of four comparable experiments (Figure S19, Table S11) are insensitive to variation of oxygen plasma exposure and resulting surface properties, with a mean analytical yield of 30 % ± 7 % and apparent stability for up to 100 h (Video S1), which is near the end of the duration of device capillarity. Smoother surfaces may decrease adsorption and increase yield.[30] Faster flow and surface chemical engineering of replicas may also mitigate adsorption. At least as importantly, however, analytical discrimination of size exclusion from surface adsorption enables the robust and repeatable analysis of colloidal nanoparticles of uncertain diameter due to unknown surface properties. The measurement throughput is scalable through both focused-ion-beam machining[56] of larger device arrays and widefield optical microscopy across multiple imaging fields. The analysis of thousands of nanoparticle images per optical micrograph, along with comprehensive calibrations, provides ample data to support our new approach to data reduction.

**Particle diameter distribution.** Size-exclusion data from four comparable experiments yield histograms of apparent diameters. We analyze these experiments individually and collectively to rigorously test our sizing accuracy, in comparison to the reference diameter histogram from transmission electron microscopy.

In a first analysis of the apparent diameter histogram, we use the prior information of the upper bound of particle diameter. An estimate of this quantity is a prerequisite for device design and is available from sample preparation processes, such as filtration of larger particles, as well as complementary analytical methods. A comparison of the upper bound of the reference diameter histogram, resulting from both empirical analysis and from the Johnson $S_U$ distribution, informs selection of an upper bound estimate of nanoparticle diameter of 134 nm (Table S4). This prior information is highly advantageous in our application, improving the accuracy of diameter measurements by the categorization of apparent diameter data into three types – true positives, true negatives, and false positives (Figure 3a-c). We define true positives to be nanoparticles inside of size-exclusion regions below the upper bound, true negatives to be nanoparticles outside of size-exclusion regions, and false positives to be nanoparticles inside of size-exclusion regions above the upper bound. Statistical identification of false positives enables a novel correction that accounts for nanoparticle attrition due to size exclusion and surface adsorption throughout the device. This correction estimates the number of spurious results at each step as the product of the false positive rate and the cumulative diameter histogram from size exclusion, and culls this number of nanoparticles from subsequent analysis with bootstrap resampling.

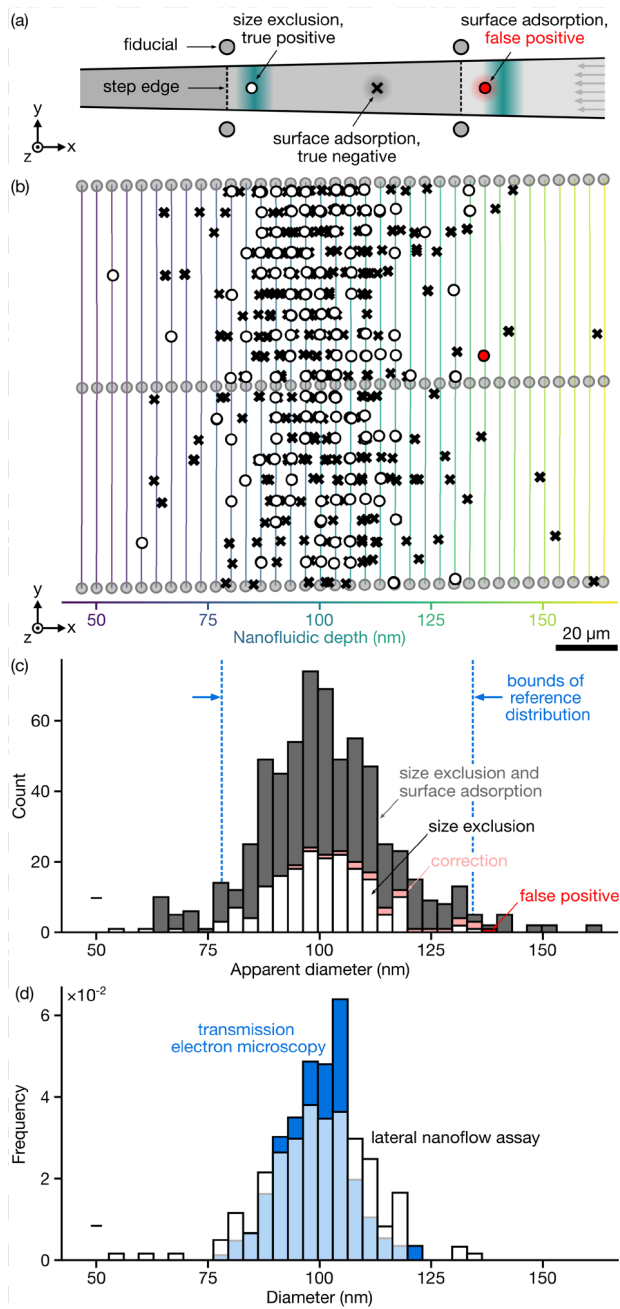

Figure 3. Diameter measurement summary. (a) Schematic showing a top view of (white circle) nanoparticles in a size-exclusion region, a range of positions in which the nanoparticle is statistically in contact with the edge of a step, and nanoparticles adsorbed to the surface of the staircase either (black ×) outside of or (red circle) inside a size-exclusion region, but beyond an acceptable upper bound of the diameter distribution. (b) Plot of nanoparticle and fiducial positions and size-exclusion regions. Localization uncertainties are smaller than data markers. (c) Histogram showing diameters apparent from nanofluidic depth for nanoparticles (gray) outside of, (white) inside of size-exclusion regions, (red) inside of size-exclusion regions but beyond an acceptable upper bound of the diameter distribution, and (pink) the correction from the upper bound. (d) Histograms from (blue) reference measurements by transmission electron microscopy and (white) the lateral nanoflow assay. The data in (b-d) correspond to a representative experiment in Table 1. Results from all experiments are in Figure S19 and Table S12. Lone bars in the lower left corners of (c) and (d) are representative uncertainties of diameter, which we plot as 95 % coverage intervals.

The false-positive rate is approximately 6 % for all four experiments, which is only a fraction of the nanoparticles that would result from a uniform probability of attractive interactions with device surfaces. Previous simulations[57] support this result, indicating that local decreases of channel depth increase flow speed to yield repulsive hydrodynamic interactions and intrinsic robustness against attractive surface interactions near step edges. Unlike previous efforts,[23] our validation of the apparent diameter histogram achieves distributional accuracy to within errors of the mean of 2.3 nm and of the standard deviation of 4.2 nm (Table 1). These errors are generally insensitive to the upper bound over a range of tens of nanometers, although a few outlying nanoparticles can significantly increase the standard deviation (Table S12, Figure S19, Figure S20). These results test sizing accuracy for a practical application with surface interactions and significant uncertainty of the upper bound of a size distribution.

In a second analysis of the apparent diameter histogram, we make use of more prior information to test the limit of accuracy of our method in the present experimental system, applying both an upper and lower bound to reject the outlying nanoparticles and analyze apparent diameters only within the 99.9 % coverage interval of the reference diameter histogram. Incorporation of a lower bound reduces errors of the mean to 0.4 nm and of the standard deviation to 2.2 nm. The larger error is comparable to the uncertainties of single nanoparticles and exceeds the uncertainties of the distribution moments by approximately 1 nm (Table 1, Table S12). This level of accuracy highlights the ability of our measurement system to measure the diameter distribution of nanoparticles to within an error of a few nanometers, despite fundamental differences between the methods of the lateral nanoflow assay and of transmission electron microscopy, and in even in the presence of size-exclusion data that is non-ideal. In comparison to our previous study,[30] which involved size exclusion of smaller nanoparticles in silica devices that are impractical to scale out for widespread application, and microscopy across a narrower imaging field with lesser effects of aberrations, we achieve comparable relative accuracy (Table S12). In this way, our present study greatly improves the reliability and extensibility of our method for practical devices and general conditions.

This sizing validation enables an evaluation of the selection of size-exclusion region width and calibration effects (Table S12), which is informative for future studies and nonintuitive in several ways. Interestingly, our measurements of diameter are robust against many of the systematic effects, due to step-edge widths that are approximately one order of magnitude wider than the mean value of the position correction, and step depths that vary linearly. However, such tolerance is an unreliable proposition in general, as the position correction moves approximately 20 % of all size-exclusion data from interstitial regions of the device to positions that are statistically within size-exclusion regions, significantly altering sampling statistics (Table S7, Table S12). This large effect would increase in devices with a greater number of narrower steps for higher resolution, across wider fields for higher throughput, and with depths that vary nonlinearly to separate complex mixtures. For measurements of intensity, the point-spread-function filter, which has a robust basis in both optical theory and control measurements, has a dramatic effect by culling a long tail of the intensity histogram (Table S7, Table S12), which is critical to the following analysis.

**Hierarchical statistical analysis.** Pooling data from our four comparable experiments yields a joint histogram of the steric diameters and fluorescence intensities of single nanoparticles. These key data are the only such data for this ubiquitous standard, revealing a fundamental structure–property relationship (Figure 4). As such, these data are valuable and worthy of a penetrating statistical analysis of fluorescence variability. To achieve the best accuracy and reliability, we analyze this joint histogram after correcting for false positives and limiting the data to within the lower and upper bounds of the reference distribution, resulting in a total count of 622 single nanoparticles. To further test the analysis, we also study variants of the joint histogram without corrections and bounds (Figure S21, Table S12, Table S13).

The joint histogram has several features. Naturally, there is a positive trend on average, with larger nanoparticles having higher mean intensities. Surprisingly, however, for single nanoparticles within the narrow range of most diameter bins, the highest intensity is more than ten times the lowest intensity. This variability of intensity vastly exceeds the effect of 1 % that is naïvely attributable to random occupation statistics of $10^4$ fluorophores per nanoparticle, according to ensemble measurements.[30]

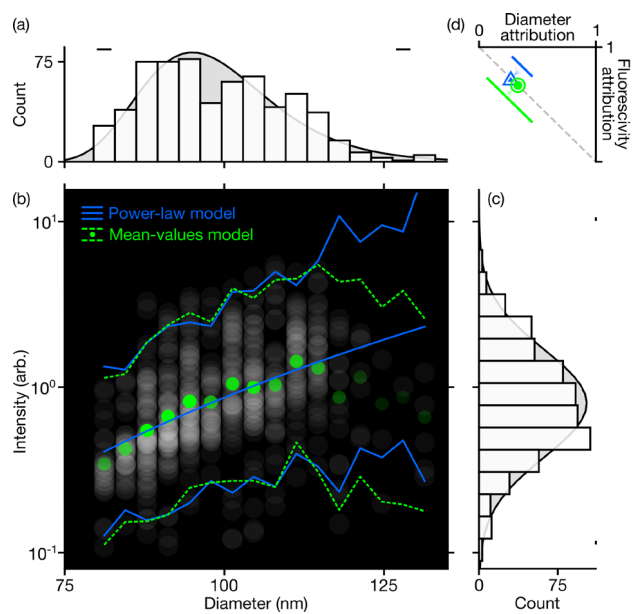

**Figure 4. Surprising nanoplastic fluorescence.** (a) Marginal histogram showing the diameters of all nanoparticles in size-exclusion regions. (Grey region) A Johnson $S_U$ distribution fits the (vertical white bars) diameter histogram data to guide the eye. (b) Two-dimensional histogram showing the fluorescence intensity of nanoparticles as a function of steric diameter. Green circles are the mean-values model. The blue line is the power-law model. Bounds are 95 % prediction intervals. (c) Marginal histogram showing the intensity of all nanoparticles in size-exclusion regions. (Grey region) A lognormal distribution fits the (horizontal white bars) intensity histogram data. Lone bars in each plot are 95% coverage intervals near the limits of either distribution. (d) Plot showing the fractions of intensity variation attributable to diameter and fluorescivity variation. Data markers are the mean values of posterior distributions. Solid lines are the major axes, or approximately 95 %, of these distributions. The minor axes are comparable to line widths. 622 nanoparticles collapse to 14 bins due to similar uncertainties for each bin. The dash gray line indicates sum to unity, with a slight distance to the data markers being attributable to uncertainty. For either model, fluorescivity and diameter attributions are independent of nanoparticle diameter. Additional results are in Figure S21, Table S13, Table S14, and Table S15.

We identify three sources of the variability of intensity and quantify the fraction of each corresponding variance to the total variance using a hierarchical statistical model.[58, 59] Accordingly, the hierarchy has three levels. The lowest level is variability due to measurement uncertainties of diameter and intensity. The relative uncertainties are small, ranging from 1.6 % to 3.2 % for diameter and 2.4 % to 3.7 % for intensity, so we expect low contributions to variability. The middle level is variability of true intensity for a constant diameter, which we assign to fluorescivity. The highest level is the variability of mean intensity across different diameters. Here, at the top of the hierarchy, we consider two possibilities. The first possibility is that the trend of mean intensity has the parametric constraint of a power-law dependence on nanoparticle diameter, which we term the power-law model. The second possibility is that each diameter bin has its own mean intensity without any parametric constraint, which we term the mean-values model.

Our purpose in considering these two possibilities is to test the power-law model. If this parametric constraint is inaccurate over the experimental range of diameters, then the mean-values model will improve accuracy by disregarding the hypothesis of a power-law dependence. Moreover, the attribution of variability from the mean-values model will be more robust to any errors of diameter because this model is insensitive to the particular values of diameter.

The power-law model estimates an exponent of 3.6 ± 0.5 (Figure 4), which is sensitive to the measurements of diameter. Application of various amounts of prior information to our corrections of the diameter data tests this sensitivity (Table S14). Without any correction, the very large apparent diameters of nanoparticles with uncharacteristically low intensities reduce the exponent to 2.1 ± 0.5. These data are influential to the fit, but are unreliable measurements of nanoparticle diameter, because they are far above the upper bound of the reference diameter histogram. As such, we can confidently remove these data from this analysis, maximizing its reliability.

An exponent of 3.6 ± 0.5 is smaller than but still comparable to that resulting from our previous analysis of the mean intensity of similar nanoparticles with a mean diameter of 48 nm and a standard deviation of 6 nm,[30] at similar wavelengths. In our previous study, a least-squares fit estimated a power-law exponent of 4.0 ± 0.2, but this simpler analysis was unable to account for uncertainty of diameter and attribute variability of intensity. Our new analysis builds confidence in the results of both studies and increases the diameter range over which a super-volumetric intensity manifests, now extending from approximately 35 nm to 125 nm. Our previous analysis of ensemble measurements of the mean intensity of similar particles with diameters ranging from tens of nanometers to tens of micrometers yielded an exponent of 3.0, indicating a potential change in scaling behavior as the particle diameter increases toward, though, and beyond the range of fluorescence wavelengths.[30]

For the data in Figure 4, fluorescivity is the dominant effect, accounting for 73 % ± 10 % of variability (Table S15). However, the diameter histogram that we measure has a standard deviation that is larger than that of the reference diameter histogram (Table S12). If our standard deviation were smaller, then the power-law exponent would increase, the attribution to variability of fluorescivity would decrease, and the attribution to variability of diameter would increase. Nonetheless, we still expect the variability of fluorescivity to dominate, because of the results from the mean-values model.

The mean-values model assumes a constant diameter within each bin and estimates a mean intensity for each diameter bin for comparison to the power-law model (Figure 4). For most of the diameter range, the mean values of intensity cluster tightly around the power-law trend, and the 95 % prediction bounds for both models are nearly identical. This pattern breaks only for the largest nanoparticles with few measurements. Both models are good descriptions of the data, because nominal 95 % prediction intervals envelop 96 % of the intensity measurements. The absence of significant disagreement builds confidence in the parametric constraint of the power law. The mean-values model also attributes the majority, 67 % ± 9 %, of the total variability in the intensity measurements to fluorescivity (Table S15).

Overall, our analysis quantitates a super-volumetric dependence of mean intensity on mean diameter and implicates fluorescivity as the dominant source of the variable intensity of single nanoparticles. For this latter aspect of the diameter–intensity relationship, both the power-law model and mean-values model assume lognormal distributions of intensities for each diameter bin, which is consistent with the experimental data and previous studies of fluorescence intensity (Table S12).[60] The lognormal distribution arises from a multiplicative product of random processes,[61] such as sorption of varying numbers of fluorophores to nanoparticles, and absorption and emission of varying numbers of photons in local molecular environments, determining the variable fluorescivity of single nanoparticles.

**Surprising nanoplastic fluorescence.** Our deeper understanding of the surprising fluorescence of a model nanoplastic has important implications, ranging from colloidal characterization and instrument calibration to hazard evaluation. To begin, previous studies have reported lognormal distributions of fluorescence intensity for polymeric nanoparticles,[60] but the causes are unknown without correlative measurements of the diameters of single nanoparticles and statistical analysis of the diameter–intensity relationship. As well, a super-volumetric intensity raises questions about basic concepts of the proportionality of fluorophore number to nanoparticle surface area or volume,[21, 30, 62] while ultrabroad and asymmetric distributions of fluorescivity can invalidate ensemble correlations attempting to infer the diameters of single nanoparticles from fluorescence intensity.[62] Unexpected diameter–intensity relationships and extreme heterogeneity are imperative to consider in the use of expected relationships as reference trends in efforts to validate emerging methods to measure single particles.[21, 23] In such measurements, the intrinsic heterogeneity of fluorescivity limits the reliability of fluorescence intensity and resulting localization precision, which can hinder instrument calibrations and yield variable uncertainty through measurements of single nanoparticles as probes, tracers, fiducials, and benchmarks.[1, 4, 36] Heterogeneous distributions of fluorophores doping the surfaces and volumes of polymeric nanoparticles can affect the detection and quantification of chemicals by ensemble[63] and single-particle[64] analysis of fluorescence intensity. Conversely, heterogeneous properties of nanoparticles can confound understanding of ensemble analysis of the sorption of toxic chemicals to polystyrene nanoparticles with surfaces terminating in carboxylic acids.[6, 8] In these ways, from optimizing the fluorescence intensity of nanoplastic products to using fluorescence intensity to elucidate the toxicology of nanoplastic byproducts, fluorescence heterogeneity is essential to understand.

## CONCLUSIONS

We develop a disposable nanofluidic device that integrates with an ordinary optical microscope to yield a lateral nanoflow assay. Our sample-in-answer-out system reduces single micrographs to joint histograms of the steric diameters and fluorescence intensities of single nanoparticles with high throughput. We study suspension hydrodynamics in complex confinement, elucidating an elegant scaling of surface forces which automates analytical separation. We approach the information limit of microscopy readout by novel calibrations, enabling accurate identification, localization, and integration of nanoparticle signals in reference to device topography across a wide field. This analysis enables a reliable analysis of size exclusion that is robust to surface adsorption, which is important for future applications to measure nanoparticles with uncertain surface properties. Defying the general expectation of dimensional instability for silicone nanofluidics,[33, 34] our scalable replicas support distributional accuracy to within errors of the mean ranging from 0.2 nm to 2.3 nm and errors of the standard deviation ranging from 2.2 nm of 4.2 nm, in comparison to a reference mean of 99.3 nm and standard deviation of 8.4 nm from transmission electron microscopy. The sizing accuracy depends on the amount of prior information that is available, requiring at least an estimate of the upper bound of the diameter distribution to implement a novel correction, and matching the relative accuracy of size exclusion in unscalable silica devices.[30] In comparison to other emerging methods,[24, 25, 65] including those that involve tracking single nanoparticles in confinement to measure hydrodynamic diameter,[20-22, 66] our method achieves record precision, accuracy, and throughput. As well, our device technology is inherently scalable, readily deployable, and operates with minimal intervention, while our data analysis obviates the need for thermometry and viscometry, trajectory analysis, and confinement correction to apply a variant of the Stokes–Einstein equation.

We apply our lateral nanoflow assay to study a ubiquitous nanoplastic standard, revealing a fundamental structure–property relationship with important implications. A hierarchical statistical analysis confirms and extends the range over which fluorescence intensity scales with steric diameter to nearly the fourth power, confounding basic concepts of surface adsorption or volume absorption of molecular adsorbates to polymeric nanoparticles. Moreover, this analysis attributes most of the intensity heterogeneity to intrinsic fluorescivity, which we define and isolate from nanoparticle diameter for the first time. These findings implicate a subwavelength effect within dielectric nanoparticles of varying diameter and a multiplicative cascade of random processes from fluorophore sorption to fluorescence emission. The resulting heterogeneity confounds any inference of dimensional and chemical properties from fluorescence intensity, warrants caution in any ensemble analysis, and limits the reliability of fluorescence intensity in measurements of single nanoparticles. In these ways, our study yields new insights into the fluorescence heterogeneity of what are among the most important nanoparticles in fluorescence spectroscopy, optical microscopy, and flow cytometry, with many applications. This surprising characterization shows the utility of our method to begin to understand structure–property relationships, even after decades of development and application of nanoparticle products that are in common use. Moreover, the unofficial standard that we test is a model of nanoplastic byproducts that sorb toxic fluorophores, such as nonylphenol, phenanthrene, and pyrene.[6-8, 26, 67, 68] The composition, size, and surfaces of environmental nanoplastics can be only more heterogeneous than that of model nanoplastics, informing future studies of both types of samples.

Other topics of interest include engineering device geometries and surface properties for sample preparation, including filtration and concentration, prior to analytical separation of nanoplastic mixtures, integration of thin films and fiducial arrays into fluidic devices to enable microscope calibration without extrinsic reference materials, and application of nanofluidic replicas to control and measure biomolecules.[69-71]

## EXPERIMENTAL METHODS

**Device fabrication.** We fabricate device molds on silicon substrates. We form microfluidic channels by photolithography and etching. We dice the substrates into chips that contain a microfluidic inlet and outlet and form thin films of silicon dioxide by thermal oxidation of the silicon chips. We complete device molds by machining arrays of nanofluidic staircase structures that connect the microfluidic inlets and outlets by focused-ion-beam milling[30] of the silica film. We replicate device molds by soft lithography[42] in two steps, first forming inverse replicas and then forming replicas in polydimethylsiloxane. We bond the silicone replicas to coverslips by bringing the surfaces into contact after brief exposure to oxygen plasma. We form submicrometer films of photoresist for optical microscopy and flatfield correction by spin-coating photoresist onto a coverslip and baking the film. We form submicrometer films of hard silicone by dilution and compression during the curing process.

**Device characterization.** We characterize the thickness of the silica film by ellipsometry. We characterize the surface topography of the silica molds, silicone inverse replicas, and silicone replicas by atomic-force microscopy.[30] We characterize the thickness of photoresist films for flatfield correction by surface profilometry.

**Fluorescent nanoparticles.** We commercially procure fluorescent polystyrene nanoparticles having carboxylate surface functionalization. The manufacturer sizes the dry nanoparticles by transmission electron microscopy, specifying a mean diameter of 99.3 nm ± 0.5 nm and a standard deviation of 8.4 nm ± 0.4 nm. We prepare a buffer system of 0.1× phosphate-buffered saline containing phosphate at a concentration of 1 mmol L$^{-1}$ and sodium chloride at a concentration of 15 mmol L$^{-1}$. We adjust the pH of the buffer to approximately 7.0 by adding hydrochloric acid and then we add nonylphenyl-polyethylene glycol at a volume fraction of 0.5 %. We disperse the nanoparticles into this buffer at a number concentration of $10^8$ mL$^{-1}$ to $10^9$ mL$^{-1}$. We prepare a sparse array of fluorescent nanoparticles on a silica coverslip. To promote adsorption, we functionalize the coverslip surface with amino groups by vapor deposition of (3-aminopropyl) triethoxysilane.

**Optical microscopy.** We record brightfield micrographs of silica molds (Figure 1e) and silicone replicas to show qualitative gradation of vertical dimensions (Figure 1f, Figure S4). We record brightfield and fluorescence micrographs of various samples for quantitative analysis. We correct micrographs that we analyze quantitatively for errors that result from nonuniform intensity of illumination and nonuniform response of the imaging sensor. Subtraction of pixel value offsets from each micrographs and normalization of the resulting pixel values by the maximum values determines a flatfield correction factor for each pixel for each micrograph type (Figure S9). We perform localization analysis of images of device fiducials and nanoparticles by open-source software.[50] We correct errors in position measursments that result from nonuniform magnification, among other optical aberrations, using an aperture array.[36] We correct errors in intensity measurements that result from optical interference in nanofluidic replicas, which causes fluorescence emission intensity to vary non-linearly with device depth.[30]

**Nanoparticle size analysis.** We infer the diameters of nanoparticles from their positions with respect to regions of size exclusion within device replicas. To compute size-exclusion regions for each step in a nanofluidic staircase, a Monte-Carlo simulation accounts for statistical variance of device dimensions, as well as fiducial and nanoparticle locations (Table S8). We filter localization data of nanoparticles by excluding localization results of nanoparticles outsiode of regions of size exclusion and by excluding localization results of any nanoparticle pairs with positions that yield distances between nanoparticles less than the sum their radii. We establish a novel correction for diameter histograms by counting nanoparticles in size-exclusion regions above an upper bound of nanoparticle diameter, in deeper regions of replicas where size exclusion should not occur. We apply this prior information to categorize nanoparticles in such deeper regions of the device as false positives. We then estimate a false positive proportion for each device as the fraction of false positives relative to the cumulative count of true positives and true negatives above the upper bound. This approach models the inherent attrition of nanoparticles that occurs as a result of size exclusion in a channel of decreasing depth and assumes that a constant portion of nanoparticles that reach a certain step are spuriously resting in size-exclusion regions due to surface adsorption. We correct true positive counts by subtracting the product of the false positive rate and the empirical cumulative diameter distribution from the histogram of size-exclusion data. The correction reduces the size-exclusion data into diameter histograms, which we analyze further to test our sizing accuracy, in comparison to the reference diameter histogram from transmission electron microscopy (Figure 3e).

**Nanoparticle intensity analysis.** We analyze the fluorescence intensity of nanoparticles sufficiently close to size-exclusion regions and of a diameter within a 99.8 % coverage interval of a reference distribution from transmission electron microscopy. We assume that the intensities follow photon statistics from shot noise and construct Poisson distributions of intensity for each nanoparticle.[52] We calibrate the intensitites for interference effects after background subtraction and flatfield correction and normalize the resulting distributions by their mean values, propagating uncertainty through Monte-Carlo simulation.[30, 47, 48]

**Hierarchical statistical analysis.** We develop two hierarchical models,[58] which we refer to as the power-law model and the mean-values model. Hierarchical models allow for explicit incorporation of multiple sources of variability. We apply noninformative improper or weakly informative proper priors to express a state of ignorance about the model parameters. For example, we apply the lognormal distribution location parameters for the mean-values model and the power-law exponent for the power-law model, before observing the data. We evaluate each model using open-source software[72, 73] to attribute intensity heterogeneity to three fractional sources – measurement uncertainty of nanoparticle diameter and intensity, variability of diameter, and variability of fluorescivity.

## AUTHOR INFORMATION

**Corresponding Author**
    **Samuel M. Stavis** – Microsystems and Nanotechnology Division, National Institute of Standards and Technology, Gaithersburg, Maryland, USA.

**Authors**
    **Kuo-Tang Liao** – Microsystems and Nanotechnology Division, National Institute of Standards and Technology, Gaithersburg, Maryland, USA; Maryland

Nanocenter, University of Maryland, College Park, Maryland, USA; Division of Therapeutic Performance 1, Office of Research Standards, Office of Generic Drugs, Center for Drug Evaluation and Research, Food and Drug Administration, Silver Spring, Maryland, USA;

**Andrew C. Madison** – Microsystems and Nanotechnology Division, National Institute of Standards and Technology, Gaithersburg, Maryland, USA;

**Adam L. Pintar** – Statistical Engineering Division, National Institute of Standards and Technology, Gaithersburg, Maryland, USA.

**B. Robert Ilic** – Microsystems and Nanotechnology Division, National Institute of Standards and Technology, Gaithersburg, Maryland, USA;

**Craig R. Copeland** – Microsystems and Nanotechnology Division, National Institute of Standards and Technology, Gaithersburg, Maryland, USA.

**Author contributions**
[ǁ]K.-T.L. and A.C.M contributed equally. S.M.S. conceived and supervised the study. S.M.S. designed the study with contributions from K.-T.L., A.C.M., C.R.C., and A.L.P. K.-T.L. and B.R.I. fabricated the devices. K.-T.L. and C.R.C. performed the experiments with contributions from A.C.M. A.C.M. and C.R.C. analyzed image data with contributions from S.M.S. A.L.P., S.M.S., C.R.C., and A.C.M. developed the calibrations and corrections. A.L.P. performed the Bayesian statistical analysis. S.M.S. and A.C.M. prepared the manuscript with contributions from all authors.

**Notes**
The authors declare no competing interests, financial or otherwise.

**ACKNOWLEDGEMENTS**
The authors acknowledge helpful comments from L. C. C. Elliott, N. Farkas, J. M. Majikes, J. C. Geist, and J. A. Liddle. K.-T. L. acknowledges support of a United States Food and Drug Administration Fiscal Year 2018 Collaborative Opportunities for Research Excellence in Science grant, and an appointment to the Oak Ridge Institute for Science and Education (ORISE) Research Participation Program at the Center for Drug Evaluation and Research (CDER), administered by the ORISE through an agreement between the U. S. Department of Energy and CDER. A. C. M. acknowledges support of a National Research Council (NRC) Research Associateship award.

# Supporting Information for
# A lateral nanoflow assay reveals nanoplastic fluorescence heterogeneity

Kuo-Tang Liao[†‡§∥], Andrew C. Madison[†∥], Adam L. Pintar[⊥], B. Robert Ilic[†], Craig R. Copeland[†] and Samuel M. Stavis[†]*

## Index

[†]Microsystems and Nanotechnology Division, National Institute of Standards and Technology, Gaithersburg, Maryland, USA. [‡]Maryland Nanocenter, University of Maryland, College Park, Maryland, USA. [§]Division of Therapeutic Performance, Office of Research Standards, Office of Generic Drugs, Center for Drug Evaluation and Research, Food and Drug Administration, Silver Spring, MD, USA. [∥]Equal contribution. [⊥]Statistical Engineering Division, National Institute of Standards and Technology, Gaithersburg, Maryland, USA. *e-mail: samuel.stavis@nist.gov. [#]We identify certain commercial equipment, instruments, and materials to specify our experimental procedure. Such identification does not imply recommendation or endorsement by the National Institute of Standards and Technology, nor does it imply that the equipment, instruments, and materials are necessarily the best available for the purpose. [♭]We report uncertainties as 95 % coverage intervals, or we note otherwise. [°]We include one insignificant figure to prevent significant rounding errors as necessary.

# Experimental Methods
## Device fabrication
### Silicon substrates
We use prime-grade, p-type silicon (100) substrates, with a resistivity from 10 Ω cm to 20 Ω cm, a thickness of 525 µm, and a diameter of 100 mm.

### Silicon chips
Our device layout for each chip consists of a single inlet and microfluidic channel to transport buffer solution and nanoparticle suspension into the deep side of nanofluidic staircase structures, and a single microfluidic channel and outlet to extract buffer solution from the shallow side. We pattern an array of microfluidic features into a silicon substrate by photolithography and inductively-coupled plasma etching. We spin-coat photoresist onto the substrate to protect it during subsequent dicing into chips with lateral dimensions of 20 mm by 20 mm for thermal oxidation.

### Silica film
We grow a silica film on silicon chips by thermal oxidation in a furnace at 1,100 °C and atmospheric pressure, with an oxygen flow rate of 1,000 mL min$^{-1}$ (1,000 sccm), and with a ratio of hydrogen to oxygen of 1.85.

### Focused-ion-beam patterning
We mill patterns in silicon substrates and in silica films using a focused beam of gallium ions. Our focused-ion-beam system is from a commercial manufacturer, and we operate the system under typical conditions and using bitmap images for pattern control.[1] The pattern parameters are in Table S1We mill devices with ion-beam currents ranging from approximately 40 pA to 800 pA across pattern areas ranging from approximately 500 µm$^2$ to 6,400 µm$^2$ within 12 h. The resulting beam profile forms step edges with submicrometer widths (Figure S17). Lower currents can form narrower step edges but require longer milling times, which can result in defective nanostructures due to drift of the focused-ion-beam system.

**Table S1. Pattern parameters**

| figure | dose (pC µm$^{-2}$) | ion-beam current (pC s$^{-1}$) | number of pixels | number of passes | pattern area (µm$^2$) |
|---|---|---|---|---|---|
| 1 (base) | 3.8 × 10$^2$ | 8.3 × 10$^1$ | 9.2 × 10$^5$ | 2,850 | 5.7 × 10$^2$ |
| 1 (steps) | 7.5 × 10$^2$ | 8.7 × 10$^1$ | 9.2 × 10$^5$ | 5,320 | 5.7 × 10$^2$ |
| 1 (fiducials) | 8.4 × 10$^2$ | 8.7 × 10$^1$ | 9.2 × 10$^5$ | 6,000 | 5.7 × 10$^2$ |
| S1a | 9.2 × 10$^1$ | 4.3 × 10$^1$ | 1.5 × 10$^7$ | 90 | 6.3 × 10$^2$ |
| S1b | 1.9 × 10$^2$ | 1.0 × 10$^2$ | 1.5 × 10$^7$ | 75 | 6.3 × 10$^2$ |
| S2 | 7.5 × 10$^1$ | 1.0 × 10$^2$ | 1.5 × 10$^7$ | 30 | 6.3 × 10$^2$ |
| S3a (base) | 2.1 × 10$^2$ | 4.9 × 10$^2$ | 1.6 × 10$^7$ | 175 | 6.4 × 10$^3$ |
| S3a (steps) | 1.1 × 10$^3$ | 4.9 × 10$^2$ | 1.6 × 10$^7$ | 910 | 6.4 × 10$^3$ |
| S3a (fiducials) | 2.4 × 10$^3$ | 4.9 × 10$^2$ | 1.2 × 10$^7$ | 2,000 | 4.9 × 10$^2$ |
| S4 | 7.5 × 10$^2$ | 7.8 × 10$^2$ | 1.6 × 10$^7$ | 350 | 5.7 × 10$^3$ |
| S7a (base)-top | 5.6 × 10$^2$ | 9.3 × 10$^1$ | 8.4 × 10$^5$ | 3,790 | 5.3 × 10$^2$ |
| S7a (steps)-top | 1.2 × 10$^3$ | 9.1 × 10$^1$ | 8.4 × 10$^5$ | 7,937 | 5.3 × 10$^2$ |
| S7a (base)-middle | 5.0 × 10$^2$ | 9.2 × 10$^1$ | 8.4 × 10$^5$ | 3,374 | 5.3 × 10$^2$ |
| S7a (steps)-middle | 1.2 × 10$^3$ | 9.3 × 10$^1$ | 8.4 × 10$^5$ | 8,166 | 5.3 × 10$^2$ |
| S7a (base)-bottom | 6.3 × 10$^2$ | 9.5 × 10$^1$ | 8.4 × 10$^5$ | 4,174 | 5.3 × 10$^2$ |
| S7a (steps)-bottom | 1.1 × 10$^3$ | 9.6 × 10$^1$ | 8.4 × 10$^5$ | 7,137 | 5.3 × 10$^2$ |
| S7a (fiducials) | 9.2 × 10$^2$ | 9.6 × 10$^1$ | 1.3 × 10$^6$ | 6,000 | 7.9 × 10$^2$ |

Dwell time per pixel is 1 µs for all cases.
Pattern parameters for Figure 1 also correspond to Figures S8, S11, S12, S14, S15c-d, S16, S17, S19a-d, and Video S1.

### Staircase structures
Our device design consists of an array of 20 staircase structures with a pitch in the y direction of 7.5 µm. Each structure has an inlet width of 2.5 µm and an outlet width of 0.5 µm, and a linear taper between the two widths along the structure length of 229.6 µm. Four bitmaps, including a base that underlies staircase steps, an inlet channel, and an outlet channel, form the pattern (Table S1).

### Fluorosilanization of silica molds
We silanize silica molds with tridecafluoro-1,1,2,2-tetrahydrooctyl-1-trichlorosilane (TFOCS) prior to each replication to ensure reliable release of silicone inverse replicas. We place the molds in a vacuum bell jar with a volume of approximately 3.6 L and deposit approximately 0.5 µL of TFOCS (Table S2) into a container adjacent to the molds. A mechanical vacuum pump with a free air displacement of approximately 50 L min$^{-1}$ evacuates the bell jar for a duration of 45 s, vaporizing the TFOCS. Further details are in our previous study.[2]

**Table S2. Replication materials**

| purpose | product name | chemical name | quantity |
|---|---|---|---|
| hard silicone | Gelest VDT-731 | (7.0 % to 8.0 % vinylmethylsiloxane)-dimethylsiloxane copolymer, trimethylsiloxy terminated | 3.4 g |
|  | Sigma-Aldrich modulator | 2,4,6,8-tetramethylcyclotetrasiloxane | 50 µL |
|  | Gelest platinum catalyst | platinum-divinyltetramethyldisiloxane | 18 µL |
|  | Gelest HMS-301 | (25 % to 35 % methylhydrosiloxane)-dimethylsiloxane copolymer, trimethylsiloxy terminated | 2 mL |
| soft silicone | Dow Corning Sylgard 184 | pre-polymer<br>curing agent | 15 g<br>1.5 g |
| silanization | United Chem. Tech. TFOCS | tridecafluoro-1,1,2,2-tetrahydrooctyl-1-trichlorosilane | 0.5 µL |

Typical recipe for silicone unless we note otherwise
We store HMS-301 at 4 °C
The quantity of Sylgard 184 is variable while the ratio of pre-polymer and curing agent is constant

**Silicone inverse replicas**

We modify our process of forming a bilayer of hard and soft silicone.[2] Briefly, this process yields a microscale film of hard silicone under a milliscale film of soft silicone. To prepare the hard silicone, we mix 3.4 g of VDT-731 with 50 µL of modulator and 18 µL of platinum catalyst (Table S2). We degas the mixture in a vacuum bell jar using a mechanical vacuum pump, add 2.0 mL of HMS-301, and gently stir. We degas the mixture again, deposit it on a silica mold with TFOCS coating, and spin-coat the mixture onto the mold at 104.7 rad s$^{-1}$ (1,000 RPM) for 45 s. After curing in an oven at 80 °C for 5 min, we place the silica mold in a petri dish with a diameter of 60 mm and pour a soft silicone mixture over the silica mold. We cure the soft silicone in an oven at 80 °C for at least 4 h and peel off the inverse silicone replica.

**Fluorosilanization of silicone inverse replicas**

We apply the same process of silanizing silica molds with TFOCS to silicone inverse replicas for mold release. However, a single process of TFOCS silanization suffices for several tens of uses.

**Silicone replicas**

We prepare silicone replicas by a process that is similar to the molding of silicone inverse replicas. One important difference is that we apply three films of hard silicone to the silicone inverse replicas, curing in an oven at 80 °C for 5 min between subsequent films, before adding soft silicone, which we cure for 4 h in a 80 °C oven. Informal tests indicate that the trilayer of hard silicone increases mechanical rigidity and decreases elastic deformation of nanofluidic structures (not shown).

**Submicrometer silicone films**

We form a submicrometer film of hard silicone by three thinning processes. First, we dilute the hard silicone with an organic solvent to reduce its viscosity. We add a mixture of 3.4 g of VDT-731, 50 µL of modulator, and 18 µL of platinum catalyst to 2.25 g of hexane and mix by vortexing. We add 1.5 mL of HMS-301 to the mixture and vortex again. After degassing the dilute mixture in a vacuum bell jar with a mechanical vacuum pump, we sieve the mixture through a filter with a pore size of 0.2 µm. Second, we deposit the dilute mixture on a silicon staircase mold after TFOCS fluorosilanization and spin-coat at a frequency of 1,047 rad s$^{-1}$ (10,000 RPM) for 5 min. Third, we press the hard silicone film by a silicon substrate on which we place a mass of 2 kg, applying a constant pressure of approximately 125 Pa during curing for at room temperature for approximately 12 h. We remove the cover wafer, transferring the submicrometer film from the substrate mold to the cover wafer and obtaining a submicrometer film of hard silicone with an inverse staircase structure (Figure S4).

**Silica coverslips**

We use microscope coverslips of ultraviolet-grade fused silica with a thickness of approximately 170 µm, a root-mean-square surface roughness of less than 0.8 nm, and a surface quality with a scratch/dig specification of 20/10.

**Device bonding**

We punch holes through silicone replicas to make inlet and outlet reservoirs at the microchannel termini. We clean a silicone replica with methanol and a fused silica coverslip with piranha solution. Exposure of silicone replicas and silica coverslips to an oxygen plasma at a pressure of 27 Pa (0.2 Torr) and a power of 18 W for less than 30 s terminates both surfaces with silanol groups. We bring the two surfaces into contact at room temperature to bond them by condensation of the silanol groups.

**Aminosilanization of silica coverslips**

We functionalize silica coverslips with amino groups by vapor deposition of (3-aminopropyl) triethoxysilane (APTES) to promote adsorption of nanoparticles to coverslip surfaces for a control measurement. During the deposition of APTES, a nozzle with a head temperature of 150 °C injects 2 mL of water with flow rate of 0.4 mL min$^{-1}$ in discrete pulses with volumes of 0.1 mL per pulse over 300 s at a frequency of 66.7 mHz into a vacuum chamber at 100 °C and at a base pressure of approximately 133 Pa (1 Torr). After one purge cycle, the chamber pressure reduces to approximately 93 Pa (0.7 Torr), and a nozzle injects 0.3 mL of APTES at a rate of 0.1 mL per pulse over 900 s (0.02 mL min$^{-1}$) into the chamber.

**Photoresist film**
We form submicrometer photoresist films for optical microscopy and flatfield correction. We spin-coat polydimethylglutarimide photoresist onto a coverslip at 104.7 rad s$^{-1}$ (1,000 RPM) for 60 s. We bake the coverslip and photoresist film first in an oven with a nitrogen flow rate of 50 mL min$^{-1}$ (50 sccm) at 180 °C for 5 min, and then in a vacuum oven at 90 °C for 3 h.

# Device characterization
We characterize the silica film into which we pattern molds by ellipsometry. The film thickness is 488 nm ± 2 nm and the index of refraction of the film is approximately 1.46 at a wavelength of 632.8 nm.

We characterize trilayer films of hard silicone after curing by surface profilometry. The thickness of each film is 22.6 μm ± 0.9 μm.

We characterize the surface topography of critical features of the silica mold, silicone inverse replica, and the silicon replica with the same atomic-force microscope and measurement parameters as in our previous study.[1] However, we image silica surfaces at a line scan rate of 1.0 Hz, whereas we image silicone surfaces at 1.5 Hz. Microfluidic channel depths are 0.48 μm ± 0.02 μm. We characterize the fidelity of pattern transfer from molds to inverse replicas and then to replicas, as well as the effects of fluorosilanization on replica surface roughness (Figures S1, S2, and S3). Table S3 presents results for test devices that share replication parameters with the nanofluidic devices but differ in channel width and step depth to facilitate access of the probe tip of the atomic-force microscope. Nanofluidic structures have lateral dimensions that exceed the lateral range of our atomic-force microscope, so we measure the surface topography of four nanofluidic channels in regions of 36 μm by 36 μm throughout the 200 μm lateral extent of the staircase structure, as well as the fused silica coverslips that seal the devices after bonding. From the micrographs, we measure nanofluidic depth, root-mean-square surface roughness, and step-edge width. The diameter of the probe tip sets a lower bound of surface roughness[3] through a steric interaction at a higher spatial frequency than that of a colloidal nanoparticle interacting with a device surface. We use these microscopy results to define statistical variables that propagate through our measurement model.

We characterize the thickness of photoresist films for flatfield correction by surface profilometry. The thickness is 210 nm ± 12 nm.

# Fluorescent nanoparticles
**Manufacturer specifications**
We use polystyrene nanoparticles that are commercially available. The manufacturer synthesizes the nanoparticles by an emulsion-polymerization process, resulting in approximately spherical particles of amorphous polystyrene. The manufacturer measures the diameters of dry nanoparticles by transmission electron-microscopy, specifying a mean diameter of 99.3 nm ± 0.5 nm and a standard deviation of 8.4 nm ± 0.4 nm (Table 1, Figure S5, Figure S19). After synthesis, the manufacturer disperses the nanoparticles into an organic solvent to sorb hydrophobic boron-dipyrromethene molecules, resulting in fluorescent nanoparticles with a peak excitation wavelength of 505 nm and a peak emission wavelength of 515 nm. The manufacturer functionalizes the fluorescent nanoparticles with carboxylate groups at a surface density of 0.07 nm$^{-2}$ ± 0.02 nm$^{-2}$. Our analysis of ensemble measurements by the nanoparticle manufacturer implies that fluorescence intensity scales volumetrically with particle diameter.[1]

**Nanoparticle suspension**
We prepare a buffer system of 0.1× phosphate-buffered saline containing phosphate at a concentration of 1 mmol L$^{-1}$ and sodium chloride at a concentration of 15 mmol L$^{-1}$. We adjust the pH of the buffer to approximately 7.0 by adding hydrochloric acid and then we add nonylphenyl-polyethylene glycol at a volume fraction of 0.5 %. The resulting buffer has an electrostatic screening distance of approximately 3 nm. We disperse the nanoparticles into this buffer at a number concentration of 10$^8$ mL$^{-1}$ to 10$^9$ mL$^{-1}$. We analyze the shape of pendant[4] drops of the nanoparticle suspension to measure the interfacial tension of the nanoparticle suspension, and we analyze the shape of sessile drops of the nanoparticle suspension to measure the contact angle of the nanoparticle suspension on fused silica coverslips and planar pieces of hard silicone less than 0.5 h after exposure to oxygen plasma.

**Sparse array**
We prepare a sparse array of fluorescent nanoparticles on a coverslip. We disperse the nanoparticles into pure water at a number concentration of approximately 10$^6$ mL$^{-1}$ to 10$^7$ mL$^{-1}$ and sonicate the suspension with an input power of approximately 50 W for approximately 8 h. We deposit the suspension onto a coverslip with APTES functionalization. After particle adsorption to the coverslip surface, we enclose the suspension on the first coverslip with a second microscope coverslip for imaging.

## Hydrodynamic transport

Soon after bonding, we pipette the nanoparticle suspension into the device inlet. The device primarily exploits a capillary force, and to a much lesser extent a hydrostatic force, to induce flow of the suspension and advect nanoparticles into the array of staircase structures (Figure S1). We estimate the resulting rates of advective and diffusive transport. Relevant hydrodynamic variables are in Table S5. For each experiment, we wait at least 6 h before illuminating the sample and recording one fluorescence micrograph of nanoparticles in the staircase structures.

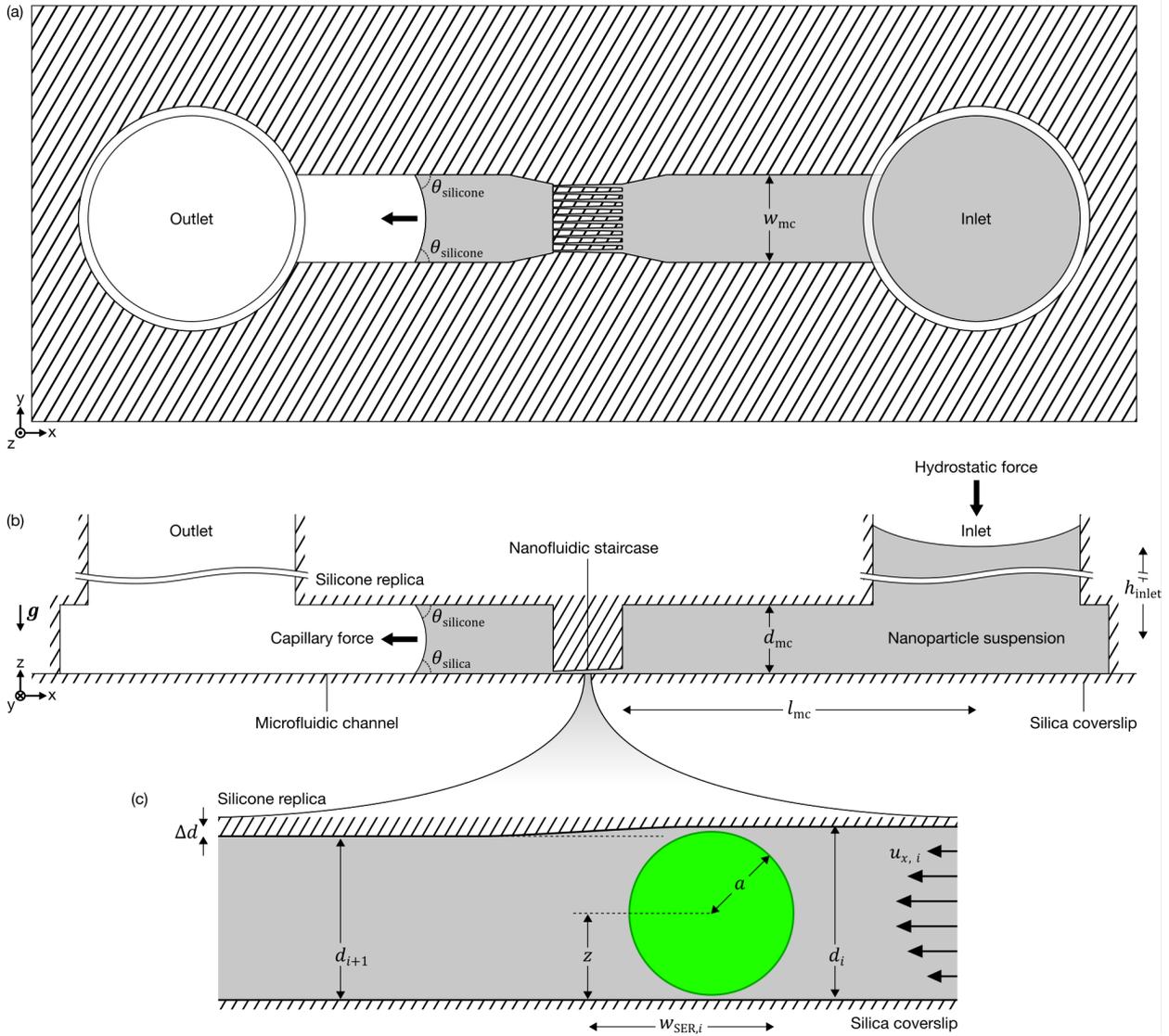

**Scheme S1. Device schematic.** Schematic showing (a) view from top, (b) view from side, and (c) detailed view from side of capillary and hydrostatic forces, which advect a suspension of fluorescent nanoparticles into a nanofluidic staircase, where size exclusion occurs. Both the width and depth of the staircase decrease from right to left in the schematic.

## Advective transport

We estimate the magnitude of the flow speed in the staircase. We calculate the pressure difference across the entire device, $\Delta p$, due to a capillary pressure difference from the air-suspension interface, $\Delta p_c$, which advances across the device as it fills, as well as a hydrostatic pressure difference, $\Delta p_g$, from the suspension column of the inlet of the device. As the device fills, the pressure difference from the capillary force exceeds that of the gravity force by a factor of $10^2$. The pressure difference, $\Delta p$, drives a volumetric flow rate, $Q$, through the entire device, with an approximation of the microfluidic channels and the nanofluidic staircase array as a series of hydraulic resistances,

$$\Delta p = \Delta p_c + \Delta p_g = \gamma \left( \frac{\cos \theta_l + \cos \theta_r}{w_{mc}} + \frac{\cos \theta_t + \cos \theta_b}{d_{mc}} \right) + \rho g h_{inlet} = QR = Q(R_{mc} + R_{sa}) \tag{S1}$$

where $\gamma$ is the surface tension of the suspension, $\theta_l = \theta_r = \theta_t = \theta_{silicone}$ are the contact angles of the suspension on the hard silicone walls at the left, right, and top of the channel of the device, $\theta_b = \theta_{silica}$ is the contact angle on the silica coverslip at the bottom of the device, $w_{mc}$ is the width of the microchannel, $d_{mc}$ is the depth of the microchannel, $\rho$ is the density of the suspension, $g$ is the acceleration of gravity, $h_{inlet}$ is the height of the fluid column above the microchannel, $R_{mc}$ is the hydraulic resistance of the microchannels, and $R_{sa}$ is the hydraulic resistance of the parallel array of staircase structures. As the device fills, the Hagen–Poiseuille law suggests that nanoparticle count in the staircase is proportional to volumetric flow rate. In this way, equation (S1) implies that nanoparticle count is proportional to inverse hydraulic resistance of the staircase. We revisit this relation in a brief study of test devices with variable depth ranges (Figure S7). Equation (S2) gives an estimate of the hydraulic resistance of fluidic channels with rectangular geometry,[5]

$$R_i = \frac{12\eta l_i}{w_i d_i^3 \left(1 - 0.630 \frac{d_i}{w_i}\right)} \tag{S2}$$

where $\eta$ is the dynamic viscosity of the suspension, $l_i$ is the length, corresponding to the $x$ direction of the $i^{th}$ segment of the channel, $w_i$ is the width corresponding to the $y$ direction of the $i^{th}$ segment of the channel, and $d_i$ is the depth, corresponding to the $z$ direction of the $i^{th}$ segment of the channel. The microfluidic channels are approximately rectangular, so we calculate $R_{mc}$ directly from equation (S2). In contrast, we calculate $R_{sa}$ as the equivalent hydraulic resistance of the parallel array of 20 staircases. In turn, we calculate the resistance of each staircase in the array, $R_s$, as a series of hydraulic resistances of each of the 36 steps,

$$R_{sa} = \left[\sum^{N_a} \frac{1}{R_s}\right]^{-1} = \frac{1}{N_a}\sum_{i=1}^{N_s} R_i = \frac{1}{N_a}\sum_{i=1}^{N_s} \frac{12\eta l_i}{w_i d_i^3 \left(1 - 0.630 \frac{d_i}{w_i}\right)} \tag{S3}$$

where $N_a$ is the number of staircases in the array and $N_s$ is the number of steps in each staircase. After solving equation (S1) for $Q$, we calculate the pressure difference across the $i^{th}$ step of the staircase, $\Delta p_i$, in terms of the hydraulic resistance at each step. We use equation (S3) to calculate hydraulic resistance, inputting a dynamic viscosity of $0.89$ mPa s $\pm$ $0.09$ mPa s and values of the channel geometry at the $i^{th}$ step in the staircase device — width, $w_i$, length, $L_i$, each with uncertainties of 25 nm and depth, $d_i$, with uncertainties that range from 2 nm to 4 nm (Table S5, Figure S6, Figure S7).[1, 6]

$$\Delta p_i = \frac{Q}{N_a} R_i \tag{S4}$$

Estimation of the volumetric flow rate and the total volume of the device allows us to approximate a lower bound on the time necessary to fill the entire device, $t_{fill}$. Estimation of the pressure difference across the device allows us to approximate the magnitude of the flow speed in the $x$ direction at each step, $u_{x,i}$, which we calculate using the analytical solution to the Navier-Stokes equation for laminar flow in a rectangular channel at small scales,[7]

$$u_{x,i}(y,z) = \frac{4d_i \Delta p_i}{\pi^3 \eta L_i} \sum_{n=1,3,5,\ldots}^{\infty} \frac{1}{n^3}\left[1 - \frac{\cosh\left(n\pi \frac{y}{d_i}\right)}{\cosh\left(n\pi \frac{w_i}{2d_i}\right)}\right]\sin\left(n\pi \frac{z}{d_i}\right). \tag{S5}$$

**Diffusive transport**

We estimate the lateral diffusivity of nanoparticles near regions of size exclusion. Our analysis involves several approximations that underestimate diffusivity, yielding an upper bound of Brenner number in the subsequent section. We begin by calculating the diffusion coefficient of a spherical nanoparticle in free suspension, far from any confining surfaces, using the Stokes-Einstein relation,[8]

$$D_0 = \frac{k_B T}{6\pi\eta a}, \tag{S6}$$

where $k_B$ is the Boltzmann constant, $T$ is absolute temperature, and $a$ is the hydrodynamic radius of the nanoparticle. Hydrodynamic interactions between the nanoparticle and the floor and ceiling of our device hinder the diffusion of nanoparticles near regions of size exclusion. To conservatively estimate an upper bound of the diffusivity of single nanoparticles in the device, we ignore hydrodynamic interactions from the Poiseuille flow, which can only reduce diffusive transport.[9-12] For simplicity, we also ignore hydrodynamic interactions between nanoparticles. A nanoparticle suspension with number concentration of $10^8$ mL$^{-1}$ to $10^9$ mL$^{-1}$ has a mean interparticle distance of approximately 10 μm. This value is a factor of $10^2$ larger than the mean diameter of the nanoparticles, allowing us to reasonably neglect hydrodynamic interactions between diffusing nanoparticles. As the nanoparticles reach their terminal positions and concentrate near size-exclusion regions, we expect the mean interparticle distance to decrease. Accordingly, we expect hydrodynamic interactions between particles to increase in magnitude and further reduce diffusivity. Considering only the hydrodynamic interactions between nanoparticles and the surfaces of the device, the diffusivity of a nanoparticle in uniaxial confinement is,

$$D_{2w\|,i}(z, a, d_i) = D_0 f_{2w\|,i}(z, a, d_i), \tag{S7}$$

where $f_{2w\|,i}(z, a, d_i)$ is the correction term from the linear superposition approximation of the diffusivity in free suspension for a single nanoparticle in uniaxial confinement between two parallel planar surfaces.[13, 14] This correction assumes that drag forces from each planar surface independently act on the nanoparticles and manifest as a superposition of corrections for a single nanoparticle near a single planar surface, $f_{1w\|}(z,a)$,

$$f_{2w\|,i}(z,a,d_i) = \left[\frac{1}{f_{1w\|}(z,a)} + \frac{1}{f_{1w\|}(d_i-z,a)} - 1\right]^{-1} \tag{S8}$$

$$f_{1w\|}(z,a) = 1 - \frac{9}{16\omega} + \frac{1}{8\omega^3} - f(\omega) \tag{S9}$$

where $\omega = z/a$ and,

$$f(\omega) = \begin{cases} \frac{15/8}{\ln(\omega-1)} - \frac{9/16}{\omega} + \frac{1/8}{\omega^3} + \exp[1.80359(\omega-1)] + 0.319037(\omega-1)^{0.2592} & \omega \leq 1.1 \\ \frac{45/256}{\omega^4} + \frac{1/16}{\omega^5} - \frac{0.22205}{\omega^6} + \frac{0.205216}{\omega^7} & \omega > 1.1 \end{cases}. \tag{S10}$$

The linear superposition approximation of the diffusivity of a single nanoparticle in uniaxial confinement between two parallel planar surfaces has limitations. In particular, as the nanoparticle diameter approaches the separation between the planar surfaces, the model may overestimate the hindrance effects of either wall, yielding estimates of the lower bound of $D_{2w\|,i}$.[14]

**Upper bounds of Brenner number**

We compare magnitudes of advective transport to our estimates of diffusive transport by calculating the upper bounds of Brenner number, $Br$, for nanoparticles near reagions of size exclusion at each step of the staircase.[15] We compute the mean magnitude of the flow speed at the $i^{th}$ step of the staircase, $\langle u_{x,i} \rangle$, which we estimate theoretically, a characteristic length scale equal to the mean width of the size exclusion regions $\langle w_{SER,i} \rangle$, and the lateral diffusivity resulting from the uniaxial confinement of nanoparticles in the staircase.

$$Br_i(z,a,d_i) = \frac{\langle u_{x,i} \rangle \langle w_{SER,i} \rangle}{D_{2w\|,i}(z,a,d_i)} \tag{S11}$$

**Device aging**

After exposure to oxygen plasma, we expect the hydrophilicity of the silicone and silica surfaces to decrease,[16, 17] reducing the magnitude of the capillary pressure difference across the air-suspension interface and slowing advection. We estimate that the contact angles of the nanoparticle suspension on either silica or hard silicone increase from approximately 0.25 rad (14 °) soon before the experiment to approximately 1.2 rad (70 °) after 6 h, when we record fluorescence micrographs of the nanoparticles in confinement. This aging effect would reduce the advection rate and Brenner numbers by a factor of approximately 3. After approximately $10^2$ h, as the hydrophobicity of the silicone replicas recovers, we expect capillarity to approach the end of its useful duration to drive hydrodynamic transport in the device.

## Microscopy standards

**Photoresist film**

We use a photoresist film on a coverslip to develop flatfield corrections. The flatness, thickness, transparency, and autofluorescence of the photoresist film enable its use as a universal artifact for each microscopy mode in this study.

**Aperture array**

We use an aperture array with a mean pitch of 5,000 nm ± 1 nm.[18] This uncertainty of 0.02 % is an estimate of the effect of placement accuracy in the fabrication of the aperture array by electron-beam lithography. Moreover, reconfiguration of the microscope system can affect the apparent value of mean pitch by up to 0.07 %.[18] Considering these sources of error and estimates of uncertainty, we take a value of 0.1 % with a uniform coverage interval as an uncertainty estimate for the aperture array pitch in this study.

## Optical microscopy

**Micrograph acquisition**

We record brightfield micrographs of molds in silica (Figure 1a) and replicas in silicone to show qualitative gradation of vertical dimensions (Figure 1b, Figure S4). For these inspection micrographs, a light-emitting diode illuminates the samples with a wavelength range of 370 nm to 700 nm. An objective lens with a nominal magnification of 50×, a numerical aperture of 0.95, and corrections for chromatic and flatfield aberrations focuses illumination and collects reflection through air immersion. A tube lens projects images onto a color charge-coupled device (CCD) camera with an on-chip pixel size of 4.54 μm.

We record brightfield and fluorescence micrographs of a variety of samples for quantitative analysis. For both types of micrographs, a light-emitting diode illuminates the samples at a peak wavelength of approximately 460 nm with a full width at half maximum of less than 27 nm. For brightfield micrographs, a beamsplitter with 50 % reflection and 50 % transmission directs light within the microscope system. For brightfield micrographs of photoresist films, a neutral-density filter attenuates the illumination intensity. For brightfield micrographs of aperture arrays, we transilluminate an empty aperture array. For fluorescence micrographs, an excitation filter with a bandpass from 430 nm to 475 nm, a dichroic beamsplitter with a transition at 495 nm, and an emission filter with a bandpass from 503 nm to 548 nm discriminate between fluorescence excitation and emission. For fluorescence micrographs of aperture arrays, we fill the aperture array with a solution of fluorescent molecules with a concentration of approximately 100 μmol L$^{-1}$ and an emission spectrum closely resembling that of the fluorescent nanoparticles. For fluorescence micrographs of nanoparticles, an exposure time of 100 ms results in sufficiently high signals for precise localization and integration. For both types of optical micrographs, an objective lens with a nominal magnification of 63×, a numerical aperture of 1.4, and corrections for chromatic and flatfield aberrations focuses illumination and collects emission, reflection, or transmission through oil immersion. A tube lens projects the image onto a complementary metal–oxide–semiconductor (CMOS) camera with 2,048 pixels by 2,048 pixels, each with an on-chip size of 6.5 μm by 6.5 μm. A mean factor of 2.0 converts from photoelectrons to analog-to-digital units, per the specification of the camera manufacturer. We operate the camera at a sensor temperature of -10 °C by thermoelectric and water cooling, without on-board correction of pixel noise, and in fast-scan mode, and we calibrate the imaging system for these parameters. For experiments in which we localize device fiducials and fluorescent nanoparticles, we record brightfield optical micrographs of device fiducials for registration immediately before we record fluorescence micrographs of nanoparticles.

The microscope systems equilibrate for at least 1 h before we record optical micrographs near the z position of best focus.

We show all optical micrographs after flatfield correction and background subtraction.

**Flatfield corrections**

To reduce errors in measurements of intensity and position that would otherwise result from nonuniform intensity of illumination and nonuniform response of the imaging sensor, we develop three independent flatfield corrections for the three imaging modes of transillumination brightfield microscopy of aperture arrays, epi-illumination brightfield microscopy of device fiducials, and epi-illumination fluorescence microscopy of aperture arrays and nanoparticles. In a previous study, we showed that the flatfield correction for our imaging sensor is approximately independent of signal intensity.[18] In this study, we record optical micrographs of photoresist films with intensity values near the middle of the sensor range. We record and average 1,000 transillumination brightfield micrographs at a single region of the film, and approximately 100 epi-illumination brightfield micrographs and approximately 100 epi-illumination fluorescence micrographs each at different regions of the films. We filter the epi-illumination brightfield micrographs and epi-illumination fluorescence micrographs by inspection to reject any micrographs with photoresist defects. Subtraction of pixel value offsets and normalization of the resulting pixel values by the maximum values determines a flatfield correction factor for each pixel for each micrograph type (Figure S9). The selection of maximum value, rather than mean value, for normalization is arbitrary and enables flatfield correction of micrographs with pixel value saturation. Flatfield correction converts pixel values from analog-to-digital units to arbitrary units with a maximum value of 65,535.

**Localization analysis**

For pixel values without saturation, subtraction of pixel value offsets and division of the resulting pixel values by the corresponding flatfield correction factors prepares images for localization analysis by open-source software.[19] We do not modify pixel values with saturation. Input settings for our CMOS camera include a root-mean-square readout noise of 1.9 electrons, a mean value of conversion factor of 2.0 analog-to-digital units per photoelectron, and an image pixel size of 100.05 nm for brightfield micrographs and 100.63 nm for fluorescence micrographs after position correction, as we describe below. We neglect the effect of flatfield correction on shot noise in maximum-likelihood estimation. Approximation of the variable readout-noise of individual pixels by their root-mean-square readout noise causes a negligible error (Figure S15). A wavelet transform approximates initial locations of fiducials and nanoparticles by applying a threshold filter with a basis spline order of three and scale of 2.0. Peak intensity thresholds of the standard deviation of the first wavelet level of each input image correspond to local maxima for neighborhoods of eight pixels. A multiple-emitter algorithm localizes nanoparticles that can be in proximity near the limit of imaging resolution. The algorithm fits symmetric Gaussian approximations of the point spread function of the microscope system, $\text{PSF}_G(x, y)$, to each emitter image by maximum-likelihood estimation, yielding measurements of the signal intensity, $I_j$, standard deviation of the Gaussian model, $\sigma_{Gj}$, x position, $x_0$, y position, $y_0$, and background signal level, $b_{Gj}$, of the of the $j^{\text{th}}$ nanoparticle.[20]

$$\text{PSF}_{Gj}(x, y) = \frac{I_j}{2\pi\sigma_{Gj}^2} \exp\left\{-\frac{(x - x_0)^2 + (y - y_0)^2}{2\sigma_{Gj}^2}\right\} + b_{Gj} \tag{S12}$$

Various fitting parameters constrain the analysis to yield reproducible results, including a fitting radius of 13 pixels, an initial standard deviation of the Gaussian model of 3.0 pixels, a maximum number of emitters per region of interest of 5, and a model selection threshold of $1 \times 10^{-6}$. The same parameters apply to all micrographs. Localization results include x and y positions, standard deviations of Gaussian models, total signal intensity, offset, and a standard deviation of the background signal for each region of interest.

**Fiducial images**

In fiducial images, some pixel values saturate due to high intensities of brightfield micrographs. We test the effect of pixel value saturation on localization accuracy by simulating and localizing symmetric Gaussian approximations of fiducial images with signal intensities that, after pixelation, approach and then exceed 65,535 arbitrary units in some pixels. We limit these pixel values to a maximum of 65,535 arbitrary units, reproducing the saturation limit of our 16-bit imaging sensor (Figure S11). We simulate 5,000 images that closely resemble the experimental images, with corresponding values of image pixel size, Gaussian standard deviation, signal-to-noise ratio without and with saturation, number of pixels that saturate, and relevant parameters of the CMOS imaging sensor including variable offset, total noise, and response.[18] We localize synthetic images of fiducials without and with saturation to assess localization accuracy (Table S6).

**Nanoparticle images**

For nanoparticle images, we reject any localization result with an intensity less than 5,000 arbitrary units and a theoretical localization precision[21] of more than 10 nm. We also reject any replicate results that coincide in position within a factor of 10 of the theoretical localization precision of single nanoparticles, retaining the result with the lowest uncertainty in the group (Table S9, Figure S12). We then filter the localization data of nanoparticles in replicas to preclude potential analysis of apparent agglomerates, which have standard deviations of the symmetric Gaussian approximation of the point spread function in excess of those of single nanoparticles (Figure S13). In a control experiment, we record epi-fluorescence micrographs of a sparse array of nanoparticles on a microscope coverslip. We fit symmetric Gaussian functions to the nanoparticle images as we describe above, compute a histogram of standard deviations, and identify the lower and upper bounds of the 95 % coverage interval of the distribution of standard deviations. Experimental values of standard deviation exceed the theoretical value[20] of approximately 92 nm due to field curvature and deviations from best focus across the imaging field. The upper and lower bounds of the histogram provide a statistical characterization of the point spread function of single nanoparticles across the imaging field, including effects of optical aberrations. We filter nanoparticles with standard deviations outside of these bounds, omitting 1,686 nanoparticles or approximately 35 % of the emitters that pass the intensity and replicate filters, which we apply prior to the analysis of size exclusion (Table S9).

**Position calibration**

We correct errors in position measurements that result from nonuniform magnification, among other optical aberrations, using an aperture array. For both brightfield and fluorescence microscopy, we record a series of micrographs of the aperture array through focus in axial increments of 10 nm, and localize each aperture in each micrograph. For fluorescence microscopy, we fill the aperture array with a fluorophore solution that has an emission spectrum that closely resembles the emission spectrum of the fluorescent nanoparticles that we measure.[1] Imaging the aperture array through focus at wavelengths matching those of the fiducials and nanoparticles allows calculation of independent correction functions at different focal positions for fiducials and nanoparticles.[18] A similarity transformation registers this set of aperture positions to those in an ideal array with a pitch of 5,000 nm, with the transformation scale factors providing mean values of image pixel size. The effects of optical aberrations, placement precision from the nanofabrication process, and fitting errors produce registration errors that have both random and systematic components, the latter showing a complex dependence on position in the imaging field for the different imaging wavelengths and positions of best focus. A linear combination of Zernike polynomials models these systematic errors, providing a function to correct positions. This correction function has an axial dependence which requires that the z positions of reference micrographs match those of the nanoparticle and fiducial position data from the nanofluidic device. For both microscopy modes, we accomplish this by selecting from either series the micrographs at best focus. The results of these position corrections are in Figure S15. After position correction of device fiducials, the three rows of 36 fiducials, totalling 108 fiducials, have a mean pitch of 5,364 nm ± 1 nm. In comparison, the nominal value of fiducial pitch is 5,360 nm. This is the first use of a standard that we fabricated previously by electron-beam lithography[18] to test the placement accuracy of a standard that we fabricate presently by focused-ion-beam milling. The root-mean-square values of position errors of 6.5 nm ± 0.9 nm in Figure S15 (e-g) indicate apparent non-linearity of the rows due to aberrations of the optical microscope system and actual non-linearity of the rows due to aberrations of the focused-ion-beam system and are consistent with a root-mean-square value of position errors parallel to the rows of fiducials of 5.8 nm ± 0.8 nm, which we calculate as the standard deivation of the Euclidean distance between neighboring fiducials in each row. This consistency builds confidence in our correction of position errors. Moreover, we test nanoparticles in isolation (Figure S13) to determine that our approximation of the readout noise of individual pixels by their root-mean-square value results in localization errors with a root-mean-square value of 0.03 nm, which is negligible in comparison to the smallest values of localization precision.

**Intensity calibration**

We correct errors in intensity measurements that result from optical interference in nanofluidic replicas, which causes fluorescence emission intensity to vary non-linearly with device depth. We calibrate this interference effect by filling the nanofluidic staircase with an aqueous solution of a fluorophore that has an emission spectrum closely resembling that of the fluorescent nanoparticles and that also has carboxylate terminal groups. Phosphate buffered saline with an ionic strength of 0.1 mmol L$^{-1}$ phosphate and 1 mol L$^{-1}$ sodium chloride reduces the electrostatic screening distance to approximately 0.3 nm[22] and minimizes any effects of electrostatic repulsion on filling uniformity. A fluorophore concentration of 4 mmol L$^{-1}$ results in high values of emission intensity in fluorescence micrographs. We measure the mean emission intensity of fluorophore solution filling each step of the staircase from regions with widths of at least 2 µm and lengths of at least 5 µm. We measure corresponding values of mean emission intensity from background just outside of the nanofluidic staircase and analyze their spatial variation due to nonuniform excitation intensity. Normalization of this function by its mean value yields a multiplicative correction for the fluorescence intensities of nanoparticles in the staircase. Subtraction of background emission intensity from fluorophore emission intensity, and division of the resulting values by the multiplicative correction, results in fluorescence intensity in arbitrary units as a function of nanofluidic depth (Figure S16). We normalize fluorescence intensity and nanofluidic depth by dividing them by their values in the center of the device, which corresponds to a nanofluidic depth of 106 nm ± 3 nm. We calculate a ratio of fluorescence intensity after normalization to nanofluidic depth after normalization as a calibration function for the subsequent analysis of single nanoparticles.

## Nanoparticle size analysis
### Reference analysis
We refer nanoparticles in size-exclusion regions to nanofluidic depths. Our device design includes fiducials at step edges of staircase structures as reference positions. Correction of fiducial and nanoparticle positions (Figure S15) reduces root-mean-square position errors in the x direction to less than 6 nm. Fitting quadratic polynomials to correct y positions of fiducials as a function of correct x positions of fiducials by damped least-squares with uniform weighting models the positions of step edges in image space, $r = (x, y)$ and approximates the distortion of fiducial placement. Uncertainties of the fit-parameters propagate into the spatial registration of the centers of each size-exclusion region. To compute size-exclusion regions for each step, a Monte-Carlo simulation accounts for variance of device dimensions, as well as fiducial and nanoparticle positions (Table S8). Summation of the nanofluidic depth, which we approximate as a uniform distribution between steps $i$ and $i + 1$, with uncertainties from the root-mean-square surface roughness of the replica and the coverslip and relevant uncertainties from atomic-force microscopy, yields the diameter, $2a_i$, of nanoparticles near the $i^{\text{th}}$ step edge,

$$d_i = 2a_i = \mathcal{U}(d_i, d_{i+1}) + \mathcal{N}(0, R_{q,r}^2) + \mathcal{N}(0, R_{q,c}^2) + \mathcal{N}(0, \sigma_{\text{calibration}, i}^2) + \mathcal{N}(0, \sigma_{\text{roughness}}^2) + \mathcal{N}(0, \sigma_{\text{flatness}}^2), \tag{S13}$$

where $\mathcal{U}$ and $\mathcal{N}$ denote uniform and normal distributions, $d_i$ and $d_{i+1}$ denote device depth from atomic force microscopy, $R_{q,r}$ and $R_{q,c}$ denote root-mean-square surface roughness of the replica and coverslip, respectively, and $\varepsilon_{\text{calibration}, i} \sim \mathcal{N}(0, \sigma_{\text{calibration}, i}^2)$ denotes a relative uncertainty of 0.5 % from calibration of the atomic-force microscope, $\varepsilon_{\text{roughness}} \sim \mathcal{N}(0, \sigma_{\text{roughness}}^2)$ is the uncertainty from the configuration of scan rate, scan resolution, and probe-tip radius,[1] and $\sigma_{\text{flatness}} \sim \mathcal{N}(0, \sigma_{\text{flatness}}^2)$, accounts for lateral flatness errors.[1]

Step edges, $\xi_i$, significantly broaden the size-exclusion regions. Atomic-force micrographs show step edges with profiles that we approximate by error functions. The widths of these error functions follow a lognormal distribution (Figure S17). These widths propagate into the spatial registration of the horizontal positions of the step edges, $x_{\xi_i}(y)$, which we model with quadratic polynomials to account for distortion of the positions of fiducials across the device,

$$x_{\xi_i} = a_{\xi_i} y^2 + b_{\xi_i} y + c_{\xi_i}, \tag{S14}$$

where $a_{\xi_i}$, $b_{\xi_i}$, and $c_{\xi_i}$ are the coefficients of the quadratic polynomials of best fit to the correct x and y positions of fiducials. The center positions of size-exclusion regions correspond to these polynomial with a horizontal offset, $x_{\text{offset}, i}$, which accounts for the lateral extents of the nanoparticle,

$$x_{\text{offset}, i} = \frac{d_i}{2} \sin\left(\cos^{-1}\left(\frac{d_{i-1}}{d_i}\right)\right) + \mathcal{N}(0, \langle \mathcal{LN}(\sigma_{\text{se}}, \mu_{\text{se}}, s_{\text{se}})\rangle), \tag{S15}$$

where $\mathcal{LN}$ denotes the lognormal distribution with shape, location, and scale parameters, $\sigma_{\text{se}}$, $\mu_{\text{se}}$, and $s_{\text{se}}$, respectively, and the bracket operator, $\langle \rangle$, denotes the expectation value of the distribution. As such, the distribution that we derive for $x_{\text{offset}, i}$ incorporates uncertainties from a comprehensive set of measurements of nanofluidic depth and the theoretical localization precision of fiducials, $\sigma_{f, i}$. Adding the horizontal offset in equation (S15) to the quadratic model of fiducial positions in equation (S14) yields the x position of the center of the $i^{\text{th}}$ size-exclusion region,

$$x_{\text{SER}, i} = a_{\xi_i} y^2 + b_{\xi_i} y + c_{\xi_i} + x_{\text{offset}, i}. \tag{S16}$$

Iterating through step edges, we refer nanoparticle locations, $r_j = (x_j, y_j)$ to nanofluidic depths, $d_i = 2a_i$, by computing the minimum Euclidian distance, $\mathfrak{D}_{ij}(y)$, between the center of the $i^{\text{th}}$ size-exclusion regions and the $j^{\text{th}}$ nanoparticle location.

$$\mathfrak{D}_{ij}(y) = \sqrt{(x_{\text{SER}, i} - x_j)^2 + (y - y_j)^2} \tag{S17}$$

To minimize $\mathfrak{D}_{ij}$, we analyze the derivative of $\mathfrak{D}_{ij}(y)$ with respect to $y$,

$$\frac{\partial \mathfrak{D}_{ij}(y)}{\partial y} = \frac{(a_{\xi_i} y^2 + b_{\xi_i} y + c_{\xi_i} + x_{\text{offset}, i} - x_j)(2a_{\xi_i} y + b_{\xi_i}) + (y - y_j)}{\mathfrak{D}_{ij}(y)}, \tag{S18}$$

which has critical points where

$$\mathfrak{D}_{ij}(y) = 0 \tag{S19}$$

and

$$(a_{\xi_i}y^2 + b_{\xi_i}y + c_{\xi_i} + x_{\text{offset}, i} - x_j)(2a_{\xi_i}y + b_{\xi_i}) + (y - y_j) = 0. \tag{S20}$$

We solve equation (S20) numerically to determine $y_{\text{SER},i}$, the y position of the size-exclusion region that minimizes $\mathfrak{D}_{ij}(y)$. Substitution of $y_{\text{SER},i}$ into equation (S17) yields the minimum Euclidean distance between center of the $i^{\text{th}}$ size-exclusion regions and the $j^{\text{th}}$ nanoparticle location.

We compare the resulting distributions of $\mathfrak{D}_{ij}$ to null distributions with a mean of 0 nm and variance that we sample with equation (S16) by subtraction. We compute quantiles of the difference between $\mathfrak{D}_{ij}$ and the null distributions for a value of 0 nm to determine if $\mathfrak{D}_{ij}$ encompasses a distance of 0 nm with respect to a 95 % coverage interval. Per this proximity criterion, we retain information from nanoparticles that are sufficiently close to size-exclusion regions for further analysis of steric diameter and fluorescence intensity.

Equation (S17) is central to our statistical model of nanofluidic size exclusion, reducing the measurement to a comparison of device geometry and nanoparticle position. Generally, atomic-force microscopy yields the device geometry, and localization microscopy yields position and calibration data, with accuracy from corrections of various optical aberrations including distortion and chromatic effects, to form a comprehensive set of dimensional parameters for our analysis. The combination of these data improves the accuracy of mapping of the positions of nanoparticles to positions within the device, where the depth of the device implies the diameter of the nanoparticle — close to a step edge. The output of equation (S17) is an empirical distribution of distances between a nanoparticle and a size-exclusion region, which enables discrimination between steric and chemical interactions between the surfaces of nanoparticles and the device.

**Steric filter**
We filter localization data of nanoparticles by excluding localization results of any nanoparticle pairs with positions that yield distances between nanoparticles less than the sum their radii. We compute distances between nanoparticles as the Euclidian distances between unique pairs of nanoparticles in size-exclusion regions, and we compare these distances to the sum of the radii of the nanoparticles. We exclude both nanoparticles of any pair with a ratio of distances between nanoparticles to the sum of the nanoparticle radii less than unity (Figure S18). This filter removes 530 nanoparticles or 17 % of the single nanoparticles after analysis of size exclusion (Table S9).

**Sizing correction**
We establish a correction factor for diameter histograms from individual experiments by counting nanoparticles in deep regions of replicas where size exclusion should not occur. This correction requires an estimate of the upper bound of nanoparticle diameters, which is a requirement for device design and is available from common methods of sample preparation such as filtration. In our study, reference measurements of nanoparticle diameter by tranmission electron microscopy inform the selection of an upper bound of nanofluidic depth. In deeper regions of a replica, steric interaction should not hinder the advection of nanoparticles, leaving surface adsorption as the nominal source of cessation of transport. We apply this prior information to categorize nanoparticles in such regions of a replica as false positives. We then estimate a false positive rate for each device as the fraction of false positives relative to the cumulative count of true positives and true negatives above the upper bound. This approach models the attrition of nanoparticles that occurs as a result of size exclusion in a staircase structure of decreasing depth and assumes that a constant portion of nanoparticles that reach a certain step are spuriously resting in size-exclusion regions due to surface adsorption. In a bootstrap resampling process, we correct true positive counts by randomly culling a number of true positives equal to the product of the false positive rate and the empirical cumulative diameter distribution. The correction reduces the true positive data into diameter histograms, which we analyze further to test our sizing accuracy, in comparison to the reference diameter histogram from transmission electron microscopy (Figure 3e).

**Reference diameter distribution**
We analyze reference measurements of nanoparticle diameter by transmission electron microscopy, computing their mean, standard deviation and cooresponding uncertainties. To select a model to guide the eye in some figures, we fit models of three different distributions to the reference diameter histograms using maximum-likelihood estimation and compare the relative goodness of the fits by computing the reduced chi-squared statistic, Akiake and Bayesian information criteria. These three models include the normal distribution, the lognormal distribution, and the Johnson $S_U$ distribution,[23] which has a probability density function for a continuous variable, $x$, of the following form,

$$f(x) = \frac{\delta}{\lambda\sqrt{2\pi}} \frac{1}{\sqrt{1 + \left(\frac{x-\xi}{\lambda}\right)^2}} \exp\left\{-\frac{1}{2}\left[\gamma + \delta \sinh^{-1}\left(\frac{x-\xi}{\lambda}\right)\right]^2\right\} \tag{S21}$$

where $\gamma$ and $\delta$ are shape parameters, $\xi$ is the location parameter, and $\lambda$ is the scale parameter. We use parametric boostrap estimation in conjunction with the models and fit parameters to estimate the distribution bounds, which we define to be the 0.1 % and 99.9 % quantiles of the model distributions. We compare these quantiles to lower and upper bounds from non-parametric boostrap estimation of the 0.1 % and 99.9 % quantiles of the empirical distribution (Figure 5, Table S4).

## Nanoparticle intensity analysis

### Fluorescence intensity measurements

The field of radiometry provides a framework for our measurements of fluorescence intensity. As well, quantitative radiometry presents issues of nomenclature[24] which motivate a clear definition of our measurands and their units. The fluorescence intensity of a nanoparticle is the radiant power emission from the nanoparticle per unit solid angle, and is equivalent to the photon flux emission from the nanoparticle per unit solid angle. Fluorescence intensity is a quantity that results from spatial and spectral integration of the fluorescence spectral radiance of the nanoparticle, $L_{fj}(\lambda)$, which is the radiant power emission from the nanoparticle, per unit wavelength, per unit solid angle, per unit area of a projection onto the imaging sensor, and is equivalent to the photon flux emission from the nanoparticle, per unit wavelength, per unit solid angle, per unit area of projection onto the imaging sensor. The fluorescence spectral radiance of the $j^{th}$ nanoparticle relates the fluorescence intensity of the nanoparticle, $I_j$, to three intrinsic optical properties – the number density of fluorophores, $N_j V_j^{-1}$, the absorption cross section of the ensemble of these fluorophores, $\sigma_j(\lambda)$, and the spectral quantum yield of the ensemble of fluorophores, $\phi_j(\lambda)$,

$$I_j = \iint_S \int_{\Delta\lambda} L_{fj}(\lambda) \, d\lambda dx dy = \iint_S \int_{\Delta\lambda} I_0 \Omega \frac{N_j}{V_j} \sigma_j(\lambda) \phi_j(\lambda) \, d\lambda dx dy, \tag{S22}$$

where $I_0$ is the incident power of the excitation, $\Omega$ is a geometric factor of the optical system, $V_j$ is the volume of the $j^{th}$ nanoparticle, $\Delta\lambda$ is the spectral bandwidth of the emission, and $S$ is the spatial extent of the region of interest for each nanoparticle image.[25]

We account for spatial nonuniformity of $I_0$, as well as variable responses of individual pixels, by our flatfield correction. We then measure a signal that is proportional to fluorescence intensity, $I_j$ of single nanoparticles. We temporally integrate over the exposure time, $\Delta t_{\exp}$, of each micrograph, spectrally integrate over the bandpass of our emission filter, and spatially integrate over the imaging region, the fluorescence spectral radiance of the images of single nanoparticles, which we model as symmetric Gaussian functions,

$$I_j \propto \int_{\Delta t_{\exp}} dt \iint_S \int_{\Delta\lambda} \frac{N_j}{V_j} \sigma_j(\lambda) \phi_j(\lambda) \, d\lambda dx dy. \tag{S23}$$

We define the fluorescivity of the ensemble of $L$ fluorophores within the $j^{th}$ nanoparticle,

$$\mathfrak{F}_j = \frac{N_j}{V_j} \sigma_j(\lambda) \phi_j(\lambda) = \frac{1}{V_j} \sum_{k=1}^{L} N_{jk} \sigma_{jk}(\lambda) \phi_{jk}(\lambda). \tag{S24}$$

In general, fluorescivity is the product of the number density, absorption cross section, and quantum of an ensemble of fluorophores. For nanoplastics, fluorescivity quantifies how a fluorophore ensemble, which can interact within the bounding surface and dielectric volume of a nanoscale particle, absorbs fluorescence excitation and yields fluorescence emission.

### Gaussian integral reliability

We compare measurements of fluorescence intensity by the two methods of Gaussian integration of nanoparticle images and direct summation of signal intensity. Both methods account for mean values of background noise to isolate signal intensity. Gaussian integration accounts for background noise with a fit parameter for a constant offset. Direct summation accounts for background noise by analysis of background noise around the perimeter of a region of interest. We fit a power-law model to the data with non-linear least-squares estimation with uniform weighting using the trust-region reflective algorithm and a smooth approximation of the absolute value of the fit residual, $\rho(z) = 2(\sqrt{1+z} - 1)$, as a loss function to establish robustness against outliers[26] (Figure S14).

### Nanoparticle fluorescence intensity

We analyze the fluorescence intensity of nanoparticles sufficiently close to size-exclusion regions and of a diameter within a 99.7 % coverage interval of the manufacturer specification. We assume that fluorescence intensities follow photon statistics from shot noise and construct Poisson distributions of intensity for each nanoparticle.[21] Uncertainties from background subtraction propagate into the following calculation for fluorescence intensity, $\hat{I}_j$, after normalization:

$$\hat{I}_j = \frac{\langle P(I_j) \rangle}{\frac{1}{M}\sum_{j=1}^{M} \langle P(I_j) \rangle} \left(\frac{1}{\hat{I}_{\text{calibration}, i}}\right), \tag{S25}$$

where $\langle P(I_j) \rangle$ is the expectation value of the Poisson distribution of the fluorescence intensity of the $j^{th}$ of $M$ nanoparticles and $\hat{I}_{\text{calibration}, i}$ is the intensity after normalization of the interference calibration for the nanofluidic depth, $\delta_i$, at the nanoparticle location,

$$\hat{I}_{\text{calibration}, i} = \left(\frac{\mathcal{N}(\hat{I}_{\text{fs}, i}, \sigma_{\text{fs}, i})}{\mathcal{N}(\hat{I}_{\text{fs}, N}, \sigma_{\text{fs}, N})}\right) \left(\frac{d_N}{d_i}\right), \tag{S26}$$

where $\hat{I}_{\text{fs}, i}$ and $\hat{I}_{\text{fs}, N}$ are respectively the mean intensities of the fluorophore solution at the $i^{th}$ and $N^{th}$ steps of the staircase, after background subtraction and flatfield correction, $\sigma_{\text{fs}, i}$ and $\sigma_{\text{fs}, N}$ are respectively the standard uncertainties of the mean intensities of the fluorophore solution at the $i^{th}$ and $N^{th}$ steps of the staircase, and $d_i$ and $d_N$ are respectively the nanofluidic depths at the $i^{th}$ and $N^{th}$ steps of the staircase, where $N$=19, the index of the central step in the staircase, which has a nanofluidic depth of 106 nm ± 3 nm.

## Bayesian statistical analysis

We develop a Bayesian statistical analysis using two hierarchical models,[27] which we refer to as the power-law model and the mean-values model. We evaluate each model using open-source software for statistical analysis.[28, 29] Hierarchical models allow for explicit incorporation of multiple sources of variability. We apply noninformative improper or weakly informative proper priors to express a state of ignorance about the model parameters before observing the data. Values of nanoparticle diameters and values of fluorescence intensity are measurements with uncertainties. Nanoparticles of similar diameters yield variable intensities, with different diameters yielding different mean values of fluorescence intensity. For clarity of the following discussion and analysis, we refer to intensity variability as intensity heterogeneity. The two models allow different attributions of intensity heterogeneity to three fractional sources – measurement uncertainty of nanoparticle diameter and intensity, variability of diameter, and variability of fluorescivity. The power-law model nests within the mean-values model, constraining the mean value of intensity to follow a power-law relationship with nanoparticle diameter. In contrast, the mean-values model does not explicitly include the diameter measurements, only the 14 diameter bins.

The power-law model is,

$$\log[\hat{y}_{ij}] \sim \mathcal{N}(y_{ij}, u_{ij}^2) \tag{S27}$$

$$\log[\hat{x}_{ij}] \sim \mathcal{N}(x_{ij}, v_{ij}^2) \tag{S28}$$

$$y_{ij} \sim \mathcal{N}(\alpha + \beta x_{ij}, \sigma_i^2). \tag{S29}$$

$$\frac{n_0 s_0^2}{\sigma_i^2} \sim \chi^2(n_0) \tag{S30}$$

$$n_0 \sim \text{half-Cauchy}(0, 5) \tag{S31}$$

In equations (S27–S31), $\hat{y}_{ij}$ is the value that we measure of the intensity of nanoparticle $j$ in diameter bin $i$, $y_{ij}$ is the true but unknown value of the intensity of nanoparticle $j$ in diameter bin $i$, $u_{ij}$ is what we assume to be the true value of the measurement uncertainty of $\hat{y}_{ij}$, $\hat{x}_{ij}$ is the diameter of nanoparticle $j$ in bin $i$, $v_{ij}$ are the measurement uncertainties of nanoparticle diameters, $\alpha$ is the intercept and $\beta$ is the slope of the power-law model, and $\sigma_i$ is an estimate of the intensity heterogeneity that is attributable to variability of fluorescivity of nanoparticles in bin $i$. Equation (S26) attracts the fluorescivity variances, the $\sigma_i^2$, toward the value $s_0^2$. This attraction is helpful for diameter bins with only a few nanoparticles or a single nanoparticle where estimation of a variance is difficult. We assign the attractant, $s_0^2$, a noninformative improper prior distribution and estimate the attractant from our measurements. The weakly informative but proper half-Cauchy distribution for $n_0$ ensures that the full posterior distribution is itself a proper probability distribution. We calculate the posterior distributions for the unknown parameters, $y_{ij}, x_{ij}, \alpha, \beta, \sigma_i, s_0$, and $n_0$. The fractions of intensity heterogeneity that are attributable to variability of nanoparticle diameter and fluorescivity follow from the definition of $R^2$ for Bayesian regression models.[30] In the power-law model, $V = [\beta^2/(N-1)]\sum_{i,j}(x_{ij} - \bar{x})^2$ where $\bar{x}$ is the sample mean of the $x_{ij}$, and $N$ is the total number of nanoparticles, which varies in our analyses (Table S13). $V$ is the product of $\beta^2$ and the sample variance of $x_{ij}$. By the power-law model, the fraction of intensity that is attributable to diameter variability is $V/[V + \sigma_i^2 + u_{ij}^2]$, and the fraction of intensity heterogeneity that is attributable to fluorescivity variability is $\sigma_i^2/[V + \sigma_i^2 + u_{ij}^2]$.

The mean-values model is,

$$\log[\hat{y}_{ij}] \sim \mathcal{N}(y_{ij}, u_{ij}^2) \tag{S32}$$

$$y_{ij} \sim \mathcal{N}(\mu_i, \sigma_i^2) \tag{S33}$$

$$\mu_i \sim \mathcal{N}(\mu, \sigma_\mu^2). \tag{S34}$$

$$\frac{n_0 s_0^2}{\sigma_i^2} \sim \chi^2(n_0) \tag{S35}$$

$$n_0 \sim \text{half-Cauchy}(0, 5) \tag{S36}$$

In equations (S32–S36), $\mu_i$ is the mean intensity of nanoparticles in bin $i$, $\mu$ is the mean intensity of all nanoparticles, and $\sigma_\mu$ is an estimate of the intensity heterogeneity that is attributable to diameter variability. We calculate the posterior distributions for the unknown parameters, $y_{ij}$, $\mu_i, \mu, \sigma_\mu, s_0$, and $n_0$. By the mean-values model, the variance of $\hat{y}_{ij}$ is $\sigma_\mu^2 + \sigma_i^2 + u_{ij}^2$. For nanoparticle $j$ in diameter bin $i$, the fraction of intensity heterogeneity that is attributable to diameter variability is $\sigma_\mu^2/[\sigma_\mu^2 + \sigma_i^2 + u_{ij}^2]$, and the fraction of intensity variability that is attributable to fluorescivity variability is $\sigma_i^2/[\sigma_\mu^2 + \sigma_i^2 + u_{ij}^2]$. We complete this analysis for four sets of data, corresponding to those without any correction from prior information of the diameter histogram, those with correction from a false positive rate, those corresponding to the lower and upper bounds of the diameter distribution, and those with corrections from both a false positive rate and lower and upper bounds of the distribution (Figure S21, Table S13, Table S14, Table S15).

## Supporting Results

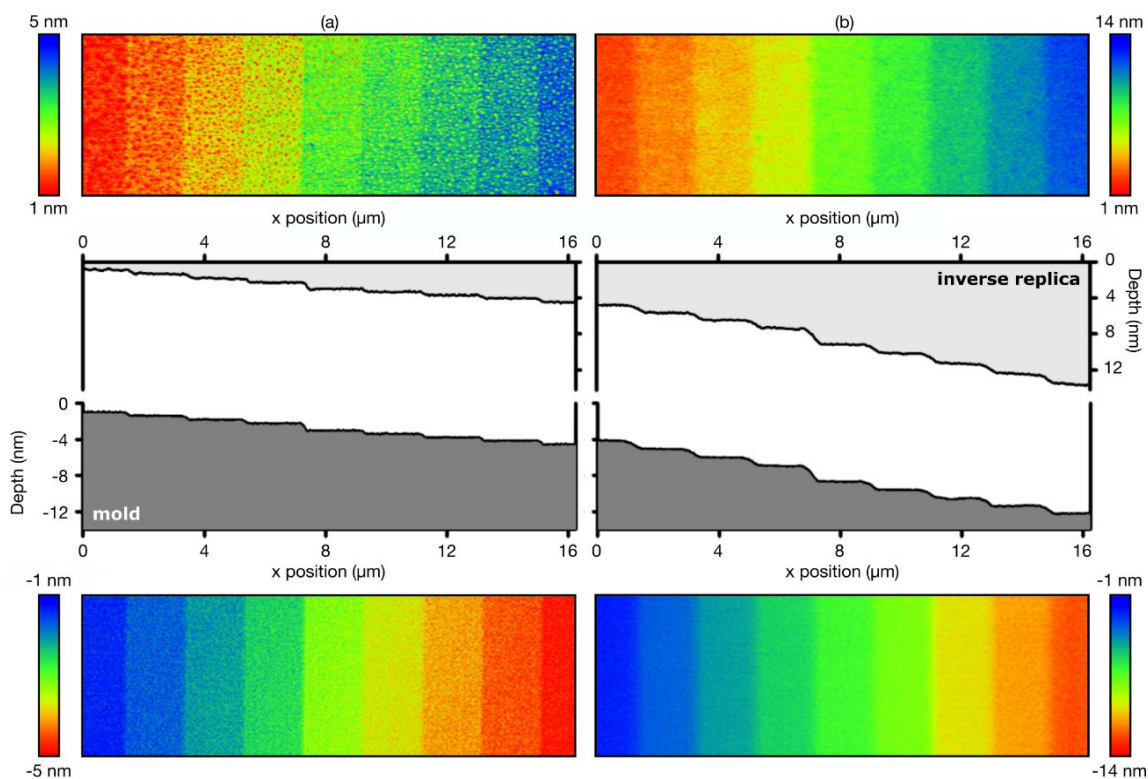

**Figure S1. Subnanometer steps.** (a-b), Atomic-force micrographs and sections showing the subnanometer fidelity of pattern transfer. The silicon mold in a has a mean step depth of 0.43 nm ± 0.13 nm and a surface roughness of 0.26 nm ± 0.01 nm. The inverse silicone replica in a has a mean step depth of 0.40 nm ± 0.14 nm and a surface roughness of 0.59 nm ± 0.06 nm. The silicon mold in (b) has a mean step depth of 0.86 nm ± 0.08 nm with a surface roughness of 0.17 nm ± 0.01 nm. The inverse silicone replica in (b) has a mean step depth increment of 0.92 nm ± 0.05 nm with a surface roughness of 0.35 nm ± 0.05 nm. Characterization of the native oxide surface of the silicon mold is prior to functionalization with tridecafluoro-1,1,2,2-tetrahydrooctyl-1-trichlorosilane (TFOCS) for mold release. The effects of mold release are evident as inverse patches in the silicone inverse replica. The TFOCS patches do not persist through a subsequent stage of pattern transfer, nor do the TFOCS patches appear on silica substrates.

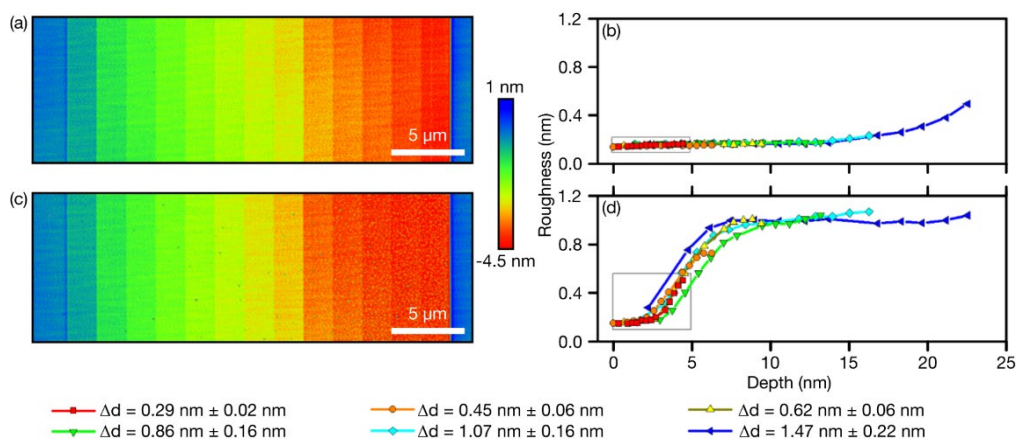

**Figure S2. Fluorosilanization effects.** (a) Atomic-force micrograph showing a staircase structure in silicon before fluorosilanization with TFOCS. (b) Root-mean-square surface roughness increases consistently with depth for staircase structures with different step depths. The gray box corresponds to the shallowest structure in (a). (c) Atomic-force micrograph showing a staircase structure in silicon after fluorosilanization with TFOCS. Patchy nanostructures form on the native oxide surface, depending on and increasing the surface roughness of the staircase structures. (d) Root-mean-square surface roughness from the TFOCS increases and then saturates at approximately 1 nm.

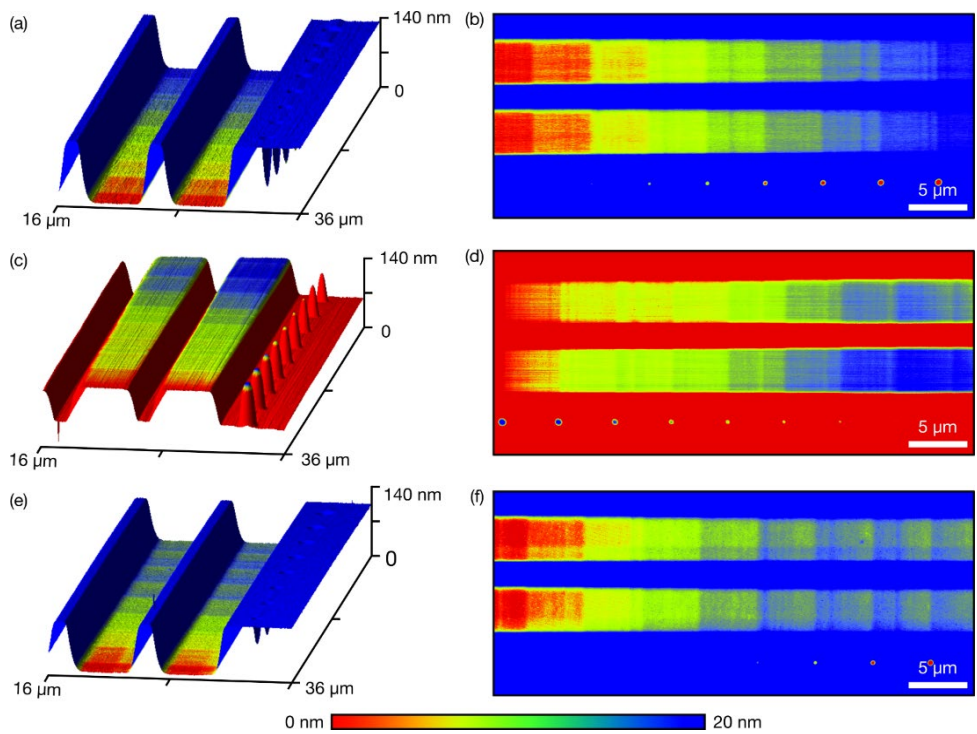

**Figure S3. Pattern transfer.** Atomic-force micrographs showing (a-b), a silica mold, (c-d), a silicone inverse replica, and (e-f), a silicone replica. To characterize replication fidelity, we use a channel width of 4.0 µm to facilitate access to the topography of all critical surfaces. In contrast, replicas in which we perform nanofluidic size exclusion have channel widths that decrease from approximately 2.5 µm to 0.5 µm as the device becomes shallower to maintain an aspect ratio that prevents channel collapse.

**Table S3. Replication fidelity**

|  | silica mold | silicone replica |
|---|---|---|
| mean increment of step depth | 1.80 nm ± 0.04 nm | 1.50 nm ± 0.10 nm |
| standard deviation of increment of step depth | 0.25 nm ± 0.06 nm | 0.65 nm ± 0.16 nm |
| surface roughness | 0.65 nm ± 0.07 nm | 0.74 nm ± 0.07 nm |

Measurements correspond to test devices that share replication parameters but differ in channel and step geometry from the experimental devices for nanofluidic size exclusion.
A nominal radius of our atomic-force microscopy probe tip of 5 nm limits the accuracy of our measurements of surface roughness, yielding an estimate of the lower bound of the actual surface roughness, which is less than approximately 1 % of the mean diameters of the nanoparticles that we measure by nanofluidic size-exclusion.

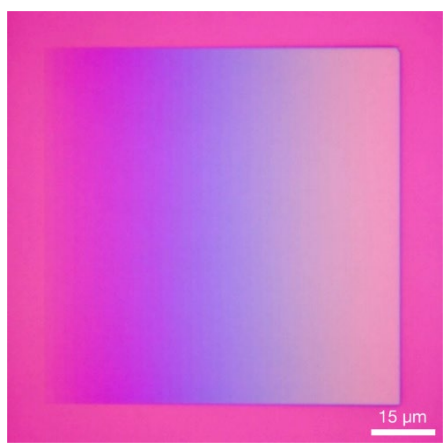

**Figure S4. Submicrometer film.** Brightfield optical micrograph showing structural colors from a film of hard silicone with a thickness of approximately 250 nm. A staircase structure rises from the zero plane of the film surface up to a height of approximately 156 nm. The root-mean-square surface roughness of the inverse silicone replica is at least 0.45 nm ± 0.07 nm, in comparison to the root-mean-square surface roughness of the mold, which is at least of 0.44 nm ± 0.07 nm.

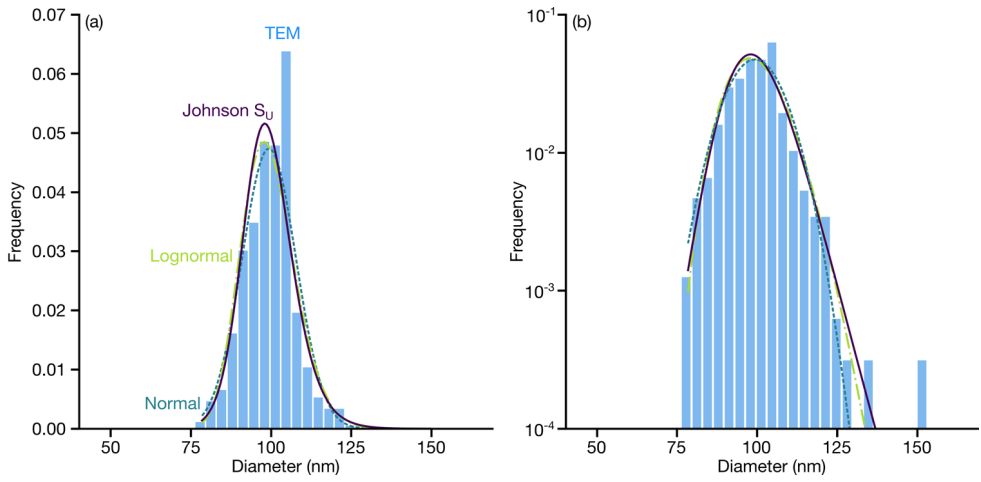

**Figure S5. Reference diameter distribution.** Histograms on (a) linear and (b) logarithmic scales showing (blue bars) reference measurements by transmission electron microscopy (TEM) with fits of various distribution models, including (solid line) a Johnson $S_U$ distribution, (dash line) a normal distribution, and (dot-dash line) a lognormal distribution. Goodness-of-fit metrics and estimates of lower and upper bounds of the reference diameter distribution are in Table S4. Fit parameters of the distribution and fits to experimental results are in Table S11.

**Table S4. Reference distribution.**

| distribution | reduced chi-squared, $\chi_\nu^2$ | Akaike information criterion | Bayesian information criterion | lower bound (nm) | upper bound (nm) |
|---|---|---|---|---|---|
| normal | $6.7 \times 10^4$ | $1.53 \times 10^3$ | $-8.31 \times 10^3$ | $74.6 \pm 4.1$ | $124.0 \pm 3.9$ |
| lognormal | $4.7 \times 10^1$ | $1.30 \times 10^3$ | $-8.31 \times 10^3$ | $78.4 \pm 2.8$ | $128.2 \pm 5.8$ |
| Johnson $S_U$ | $8.6 \times 10^0$ | $1.23 \times 10^3$ | $-8.30 \times 10^3$ | $75.5 \pm 4.8$ | $131.1 \pm 8.2$ |
| empirical | – | – | – | 78.1 | 134.3 |

We use parametric bootstrap estimation for the normal, lognormal, and Johnson $S_U$ distributions and non-parametric bootstrap estimation for the empirical distribution.
We note that non-parametric bootstrap is non-ideal for estimation of extreme quantiles and we omit uncertainties from such estimation.

**Table S5. Hydrodynamic parameters**

| parameter | symbol | value | units |
|---|---|---|---|
| height of inlet and outlet | $h_{inlet}$ | 1 | cm |
| radius of inlet and outlet | $r_{inlet}$ | 250 | μm |
| length of microfluidic channel | $l_{mc}$ | 12.7 | mm |
| width of microfluidic channel | $w_{mc}$ | 350 | μm |
| depth of microfluidic channel | $d_{mc}$ | 0.48 | μm |
| length of the entire staircase | $l_s$ | 230 | μm |
| length of each step of the staircase | $l_i$ | 5.36 | μm |
| minimum width of the staircase | $w_{i,min}$ | 0.5 | μm |
| maximum width of the staircase | $w_{i,max}$ | 2.5 | μm |
| minimum depth of the staircase | $d_{i,min}$ | 45 | nm |
| maximum depth of the staircase | $d_{i,max}$ | 165 | nm |
| number of staircases in the array | $N_a$ | 20 | – |
| number of steps in the staircase | $N_s$ | 36 | – |
| volume of entire device | $V_{device}$ | $2\times10^8$ | μm$^3$ |
| nanoparticle radius | $a$ | 45 to 165 | nm |
| surface-to-surface separation | $s_s$ | 0.1 to 10 | nm |
| surface tension of nanoparticle suspension | $\gamma$ | 37 | mN m$^{-1}$ |
| contact angle of nanoparticle suspension on fused silica | $\theta_{silica}$ | 0.22 (13) | rad (°) |
| contact angle of nanoparticle suspension on hard silicone | $\theta_{silicone}$ | 0.25 (14) | rad (°) |
| density of nanoparticle suspension | $\rho$ | 990 | kg m$^{-3}$ |
| dynamic viscosity of nanoparticle suspension | $\eta$ | 0.89 | mPa s |
| absolute temperature | $T$ | 300 | K |
| pressure difference across staircase | $\Delta p$ | $1.5\times10^2$ | kPa |
| pressure difference due to capillarity | $\Delta p_c$ | $1.5\times10^2$ | kPa |
| pressure difference due to fluid column | $\Delta p_g$ | $1\times10^{-1}$ | kPa |
| pressure difference across single step | $\Delta p_i$ | 0.135 to 33.8 | kPa |
| volumetric flow rate | $Q$ | 540 | μm$^3$ s$^{-1}$ |
| hydraulic resistance of staircase | $R$ | $2.8\times10^{-4}$ | kg μm$^{-4}$ s$^{-1}$ |
| hydraulic resistance of single step | $R_i$ | $5.0\times10^{-6}$ to $1.2\times10^{-3}$ | kg μm$^{-4}$ s$^{-1}$ |
| lower bound of filling time of the device | $t_{fill}$ | 100 | h |
| mean magnitude of flow speed in x direction | $\langle u_{x,i} \rangle$ | 0.07 to 1.2 | mm s$^{-1}$ |
| mean width of size exclusion regions | $\langle w_{SER,i} \rangle$ | 560 | nm |
| lateral diffusivity of nanoparticles | $D_{2w||,i}$ | 0.6 to 3.5 | μm$^2$ s$^{-1}$ |
| Brenner number | Br | 60 to 200 | – |

We report ranges of diffusivity and Brenner number assuming a separation of 1 nm. More detail is in Figure S6.
Quantities are approximate.

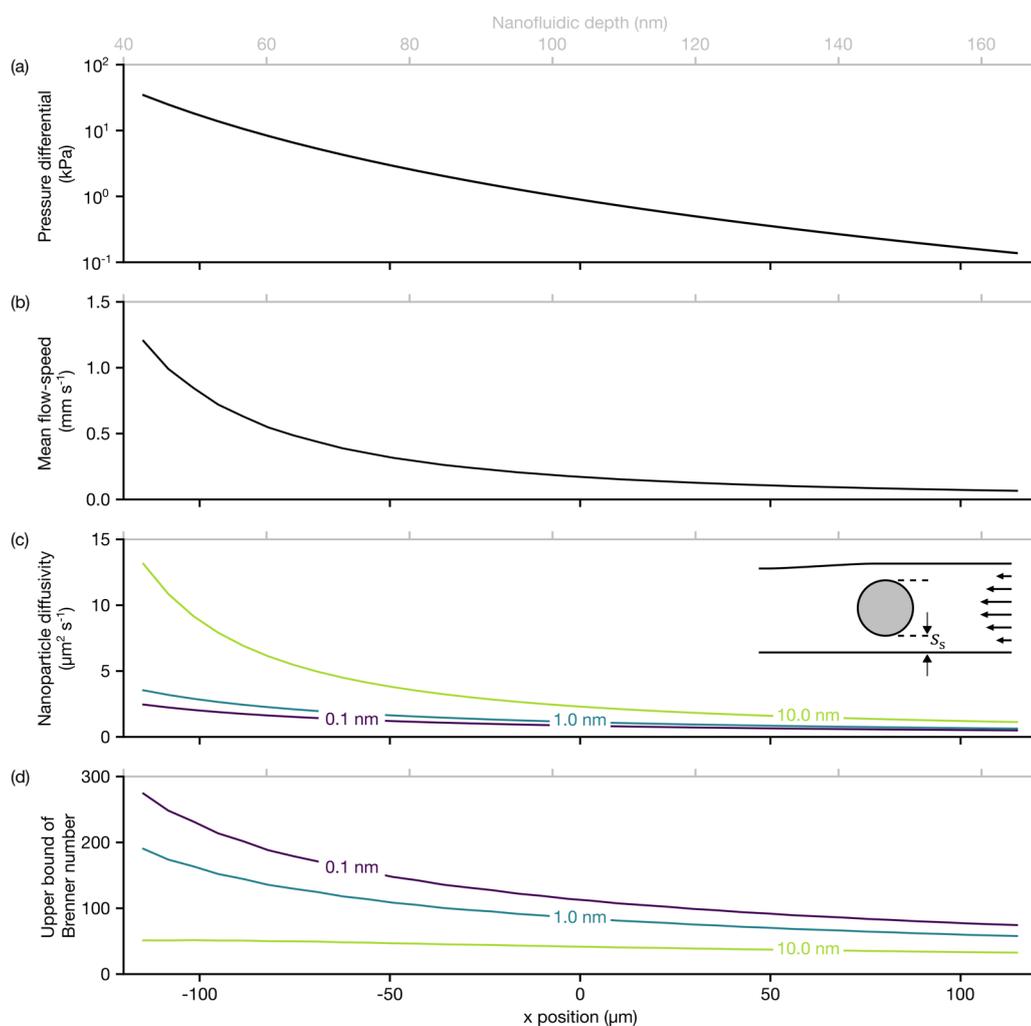

**Figure S6. Device hydrodynamics.** Plots showing theoretical values of (a) pressure difference, (b) mean magnitude of flow speed in the x direction, (c) nanoparticle diffusivity, and (d) upper bounds of Brenner numbers as a function of x position and nanofluidic depth across the nanofluidic staircase less than 0.5 h after exposure to oxygen plasma. The curves in (a-c) are upper bounds at the onset of the experiment. The curves in (c) and (d) correspond to different values of separation, $s_s$, between nanoparticles and the top and bottom of the device as we indicate in the inset of (c). Table S5 summarizes the hydrodynamic parameters that we use to estimate the rates of advective and diffusive transport of nanoparticles in the staircase.

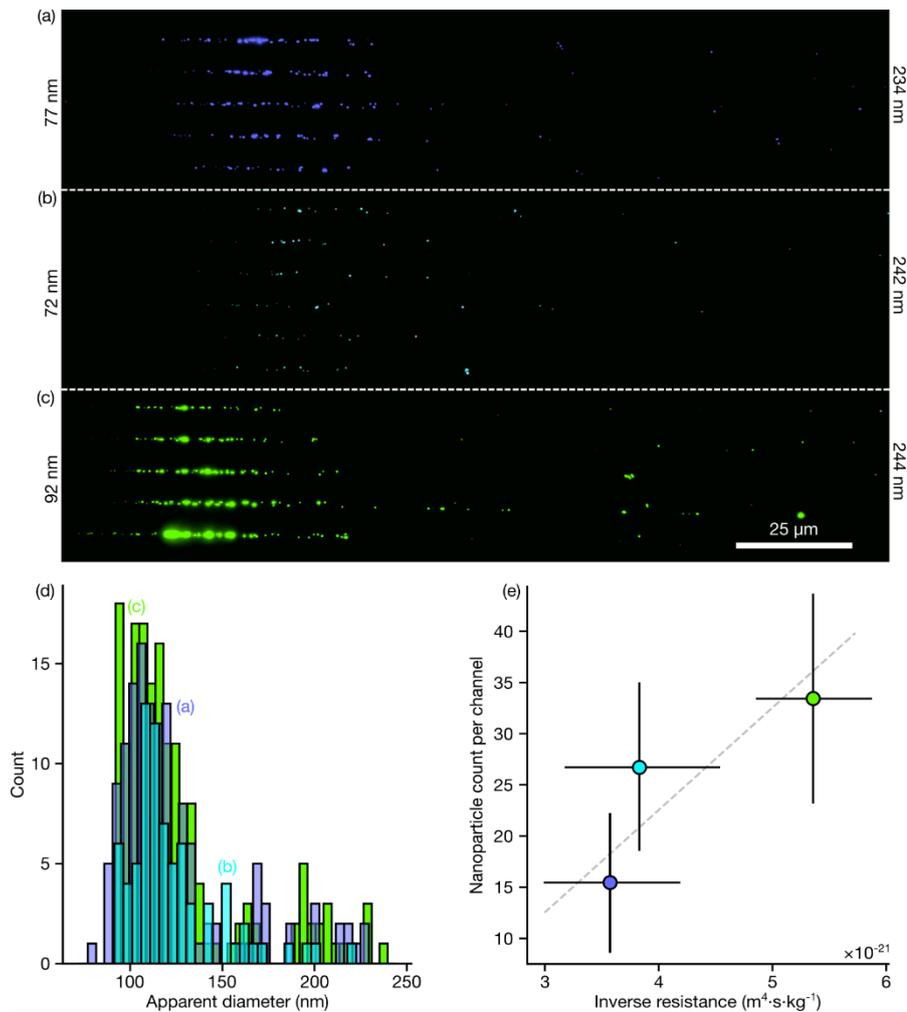

**Figure S7. Hydraulic resistance.** (a-c) False color fluorescence micrograph showing the size exclusion of fluorescent nanoparticles in three parallel arrays of nanofluidic staircase structures with variable depth ranges. (a) The top channels (blue) range in depth from 77 nm ± 3 nm to 234 nm ± 4 nm, (b), The middle channels (cyan) range in depth from 72 nm ± 3 nm to 242 nm ± 3 nm, and (c), the bottom channels (green) range in depth from 92 nm ± 2 nm to 244 nm ± 2 nm. The values on the left and right sides of (a-c) correspond with the minimum and maximum depths of each array. In a rough reduction of data for only these initial results, we omit the subsequent corrections of position and apparent diameter. (d) Histogram showing apparent diameters of nanoparticles in the top (blue), middle (cyan), and bottom (green) arrays. (e) Plot showing nanoparticle count increasing with the inverse hydraulic resistance of the arrays. We normalize particle count by the number of channels in each array. Black bars are 95 % coverage intervals. The dash line is a fit of a proportional model from equation (S1) to the data by the method of damped least squares.

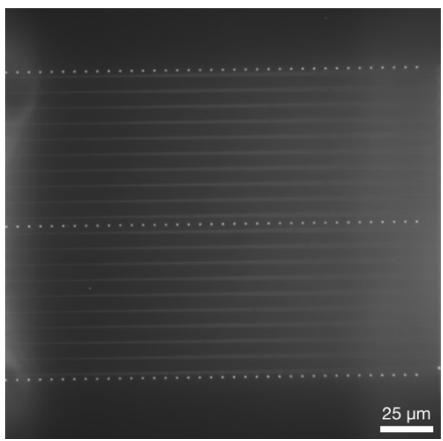

**Figure S8. Device fiducials.** Brightfield optical micrograph showing a nanofluidic device under epi-illumination after filling. Three rows of circular features are evident. The first column of device fiducials at the right edge of the micrograph marks the inlet of the staircase device, where the nanofluidic depth is greatest. All other columns of fiducials, toward the left of the micrograph, mark the 36 step edges of shallower staircase structures. In each row, the Euclidean distance between fiducials is approximately 5.36 μm.

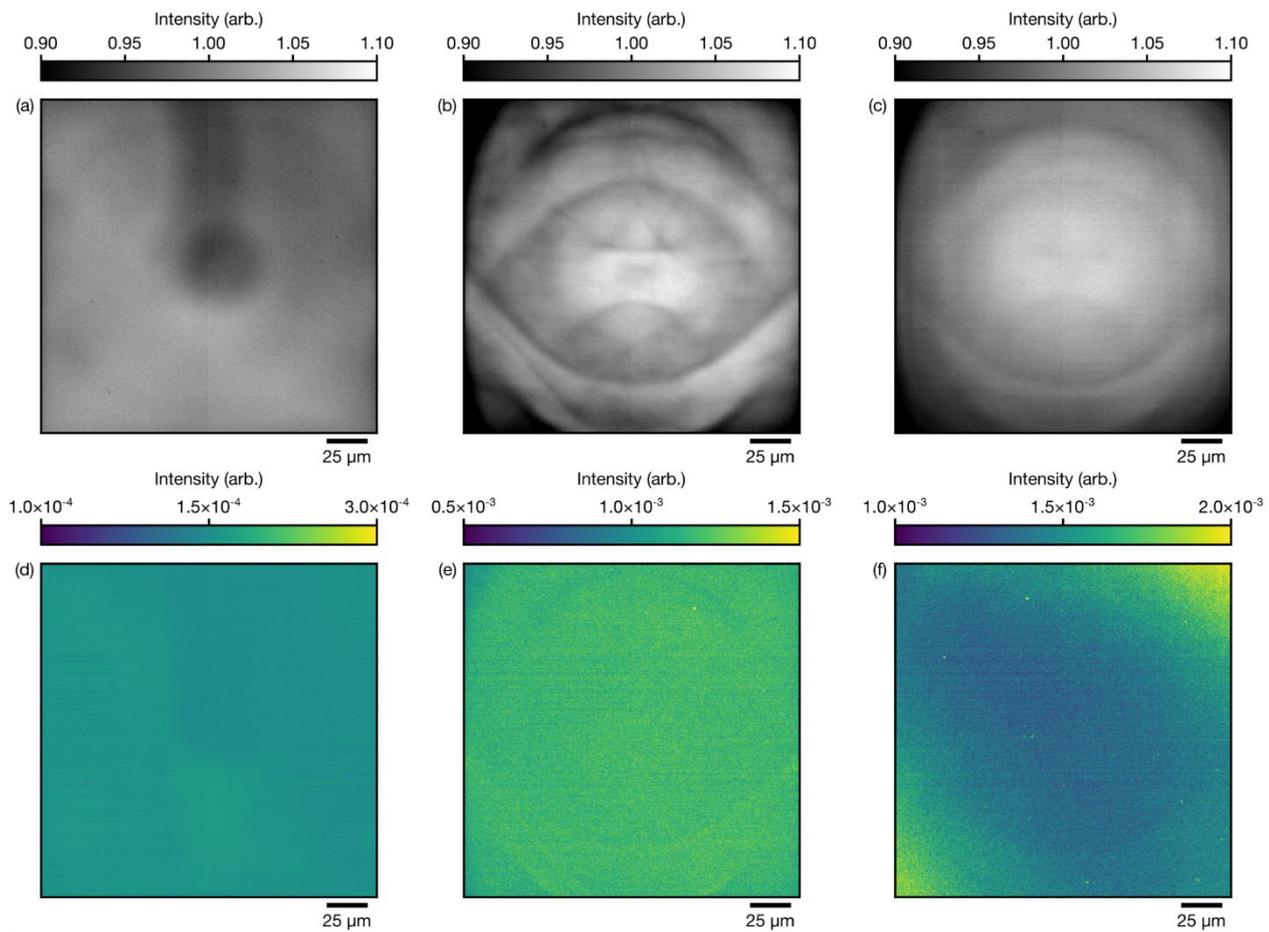

**Figure S9. Flatfield corrections.** (a, b, c) Plots showing illumination fields for (a) transillumination brightfield, (b) epi-illumination brightfield, and (c) and epi-illumination fluorescence micrographs. (d, e, f) Plots showing standard uncertainties of mean values in (a), (b), and (c), respectively. We neglect these small uncertainties. The fluorescence intensity of the photoresist film is stable within uncertainty during the measurement (not shown). The thickness of the photoresist film that we image to develop flatfield corrections is 0.2 μm, just exceeding the deepest regions of the nanofluidic devices and within the nominal depth of field of the imaging system under any illumination condition.

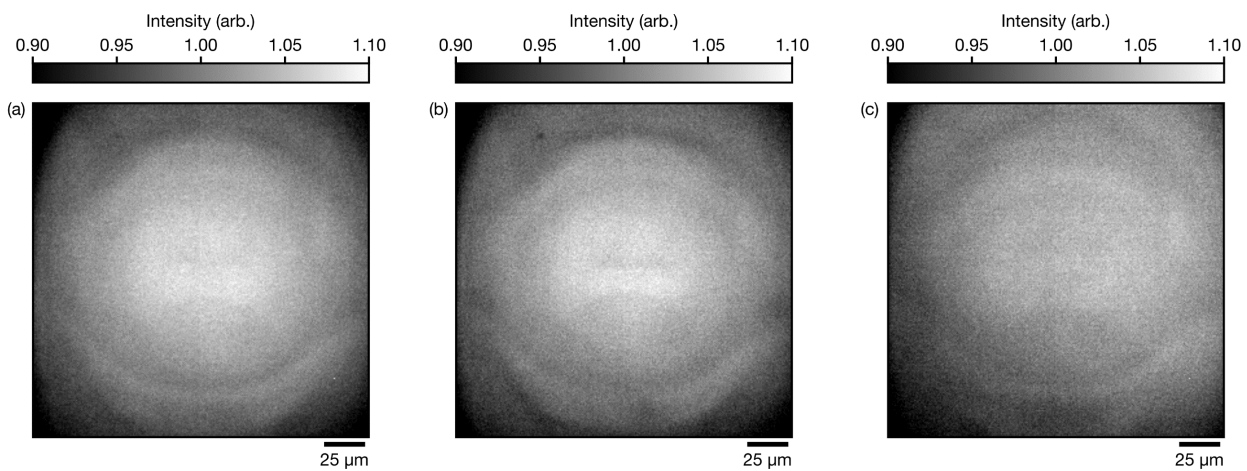

**Figure S10. Illumination fields.** Plots showing illumination fields for epi-illumination fluorescence micrographs from photoresist films with thicknesses of approximately (a) 0.2 μm, (b) 1.2 μm, and (c) 5.0 μm. The focal plane of each image corresponds approximately to the center of the film thickness.

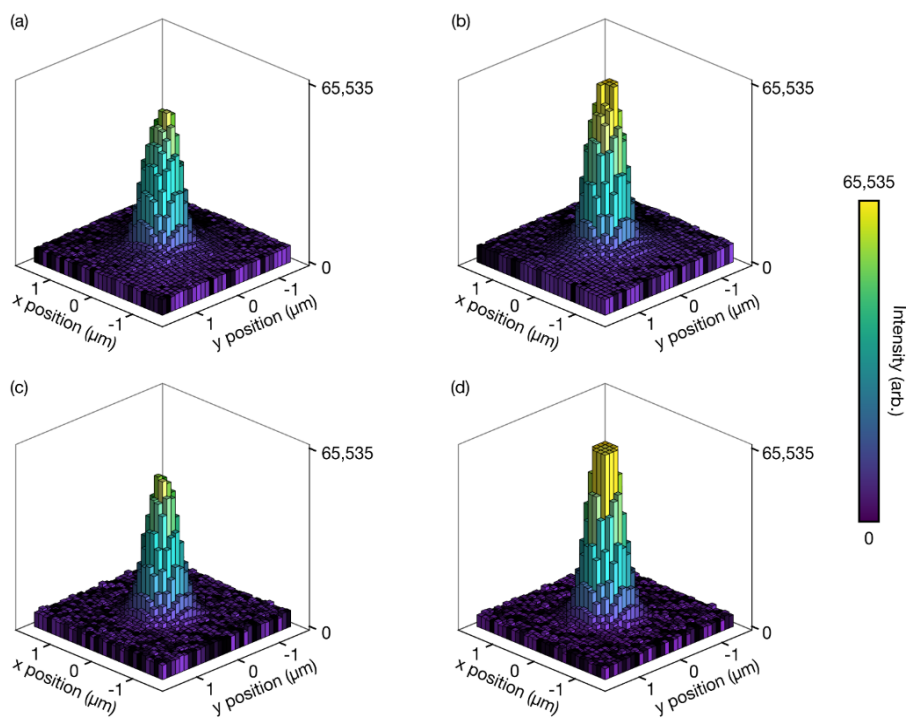

**Figure S11. Fiducial localization.** (a), (b) Surface plots showing representative brightfield micrographs of device fiducials. The signal intensity varies with illumination intensity across the imaging field. As a result, after pixelation, pixel values can be (a) below or (b) above the saturation limit of the 16-bit imaging sensor of 65,535 arbitrary units after flatfield correction. (c), (d) Plots showing synthetic images corresponding to (a), (b). Table S6 shows the corresponding comparison of the mean value of position estimates and the true position for signals below and above the saturation limit of an imaging sensor.

**Table S6. Pixel saturation**

| maximum intensity (arb.) | mean error in x-position (nm) | standard deviation of error in x-position (nm) | mean error in y-position (nm) | standard deviation of error in y-position (nm) |
|---|---|---|---|---|
| < 65,535 | -0.38 ± 0.01 | 0.39 ± 0.01 | 0.53 ± 0.01 | 0.39 ± 0.01 |
| 65,535 | -0.66 ± 0.01 | 0.31 ± 0.01 | 0.80 ± 0.01 | 0.31 ± 0.01 |

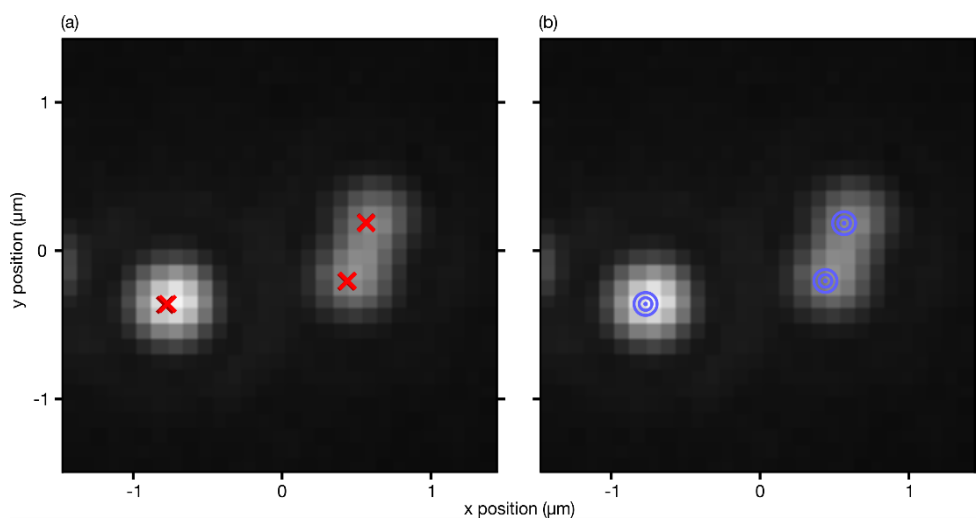

**Figure S12. Nanoparticle localization.** (a-b), Fluorescence micrographs showing representative images of nanoparticles (a), before and (b), after applying a filter to reject replicate localization results within a factor of 10 of the theoretical localization precision of single nanoparticles. (a), (red crosses) Localization of what we assume is a single nanoparticle on the left and two nanoparticles in proximity on the right yields six total positions, which reduce after filtering to (b), (blue roundels) three positions. This filter retains the localization result with the smallest localization uncertainty and rejects other localization results within each group of replicates.

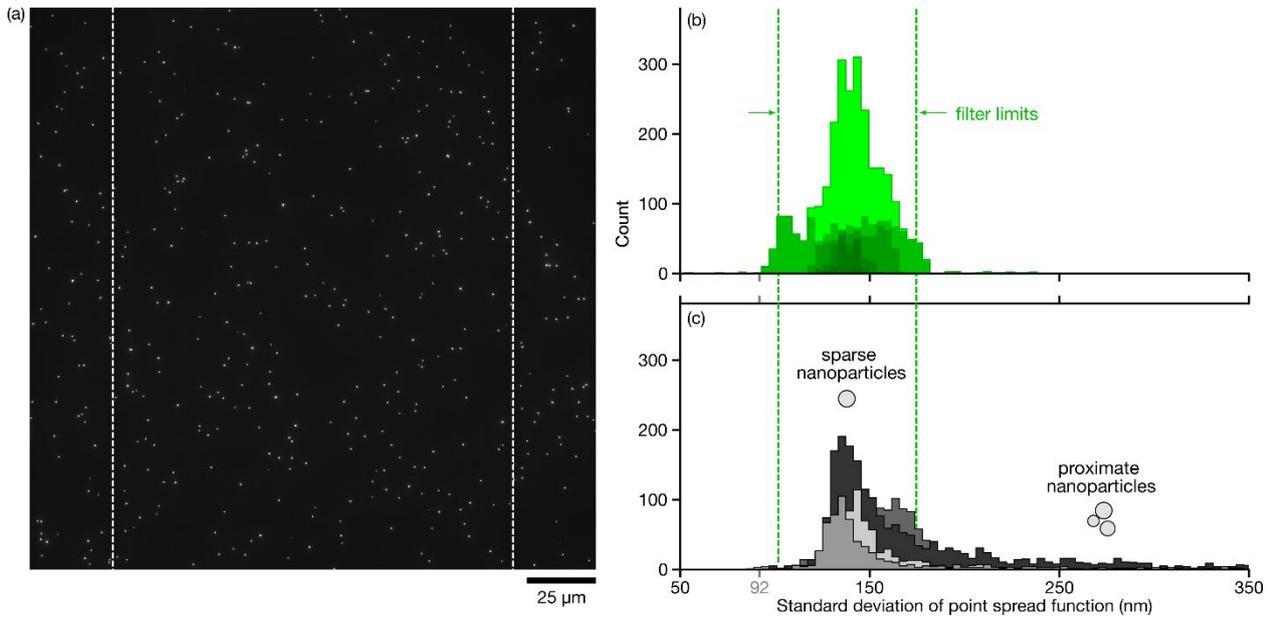

**Figure S13. Point-spread-function filter.** (a), Fluorescence micrograph showing representative images of sparse nanoparticles on a microscope coverslip in a control measurement. We determine the standard deviations of a symetrric Gaussian approximation of the point spread functions of nanoparticles between the white dash lines to match the imaging field of experiments in nanofluidic devices. (b), Histogram showing standard deviations from (white) the sum of eight control measurements such as in (a). (Gray region) A 95 % coverage interval of the standard deviation of single nanoparticles, extending from 102 nm to 175 nm, filters emitters with larger standard deviations in nanofluidic devices corresponding to multiple nanoparticles in proximity. (c) Histogram showing standard deviations from four comparable experiments in nanofluidic devices. (Gray tick mark) The theoretical value of the standard deviation of a symmetric Gaussian approximation of the point spread function at the peak emission wavelength is approximately 92 nm.[20]

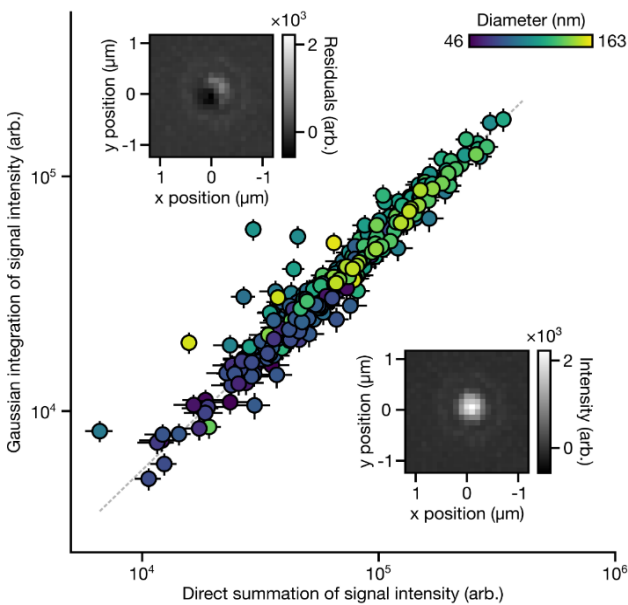

**Figure S14. Intensity measurements.** Scatter plot showing the reliable proportionality of Gaussian integration and direct summation of signal intensity in nanoparticle images. For this analysis, we select what appear to be single nanoparticles in isolation within a region of interest of 10 μm by 10 μm, avoiding any potential errors from nanoparticle images in close proximity. Fitting a (gray dash line) power-law model to the data results in a coefficient of 0.7 arb. ± 0.1 arb. and an exponent of 0.98 ± 0.01, corresponding to a mean slope of 0.5 ± 0.1 with a lower chi-square statistic, $\chi_\nu^2$, of 1.7. The resulting variability is consistent with the intensity heterogeneity values in Figure 4 of the main text. Insets: (top left) Residuals from a Gaussian model fit to a nanoparticle image and (bottom right) signal intensity of the same image.

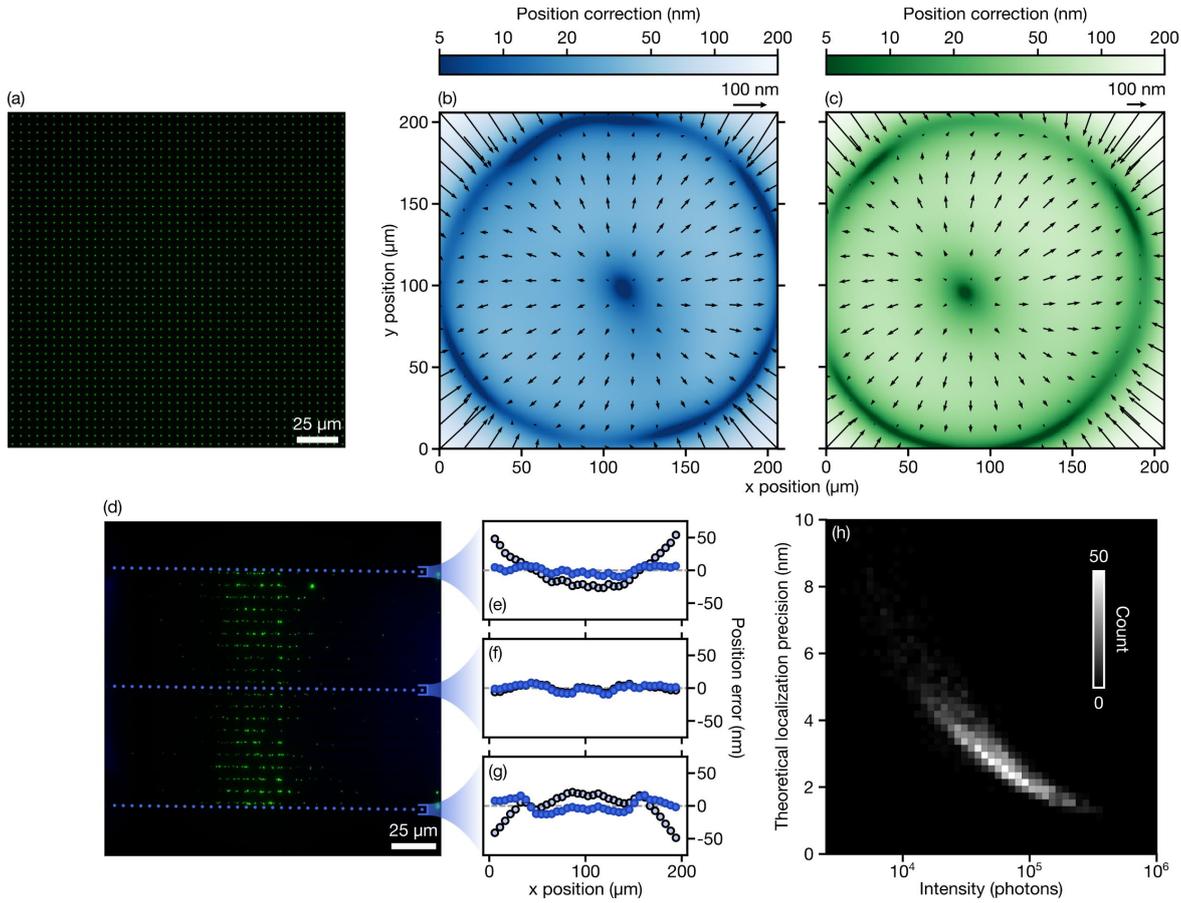

**Figure S15. Localization microscopy.** (a) (False green) Fluorescence micrograph showing an aperture array containing an aqueous solution of fluorophores with emission spectra that closely resemble the emission spectrum of the fluorescent nanoparticles that we measure. (b-c) Vector plots on a linear scale and color maps on a logarithmic scale showing independent corrections to positions of (b), fiducials and (c), nanoparticles. (d) Composite (false blue) brightfield and (false green) fluorescence micrograph showing device fiducials and fluorescent nanoparticles. (e-g) Scatter plots showing distance errors perpindicular to the row of fiducials resulting from linear fits of the three rows of fiducial positions (white circles) before correction, showing distortion of both optical micrographs and fiducial positions, and (blue circles) after correction, showing distortion of the fiducial positions only. (h) Two-dimensional histogram showing theoretical localization precision[21] of nanoparticles as a function of signal intensity among other localization parameters before flatfield correction.

**Table S7. Calibration effects**

| | experiment 1 | | experiment 2 | | experiment 3 | | experiment 4 | | all experiments | | percent change (%) | |
|---|---|---|---|---|---|---|---|---|---|---|---|---|
| calibration | $N_{ser}$ | $N_{total}$ | $N_{ser}$ | $N_{total}$ | $N_{ser}$ | $N_{total}$ | $N_{ser}$ | $N_{total}$ | $N_{ser}$ | $N_{total}$ | $N_{ser}$ | analytical yield |
| no calibration | 171 | 624 | 102 | 513 | 201 | 741 | 356 | 1503 | 830 | 3,381 | 0.0 | 0.0 |
| flatfield | 163 | 595 | 108 | 497 | 196 | 759 | 361 | 1503 | 828 | 3,354 | -0.2 | 0.6 |
| point spread function | 157 | 563 | 99 | 471 | 146 | 576 | 264 | 1165 | 666 | 2,775 | -19.8 | -2.2 |
| position | 212 | 622 | 136 | 511 | 227 | 735 | 413 | 1500 | 988 | 3,368 | 19.0 | 19.5 |
| interference | 171 | 624 | 103 | 513 | 201 | 741 | 354 | 1503 | 829 | 3,381 | -0.1 | -0.1 |
| all calibrations | 186 | 540 | 126 | 460 | 179 | 581 | 310 | 1180 | 801 | 2,761 | -3.5 | 18.2 |

We abbreviate the number of emitters inside of size-exclusion regions as $N_{ser}$.
We abbreviate the total number of emitters both inside and outside of size-exclusion regions as $N_{total}$.
Emitter count may vary by a few counts with respect to particle numbers in Tables S9, S12, and S13 due to uncertainty propagation by bootstrap resampling.
We calculate percent changes relative to the total for all experiments without calibration.
We calculate analytical yield as the fraction, $N_{ser} N_{total}^{-1}$.

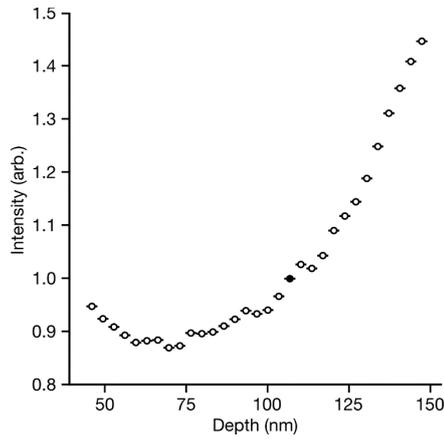

**Figure S16. Intensity calibration.** Plot showing nonmonotonic variation of fluorescence intensity due to optical interference as a function of nanofluidic replica depth. Normalization of intensity values is with respect to the intensity at the (black circle) 19$^{th}$ step of the device, which has a depth of 106 nm ± 3 nm. Horizontal bars are increments of nanofluidic depth. 95 % coverage intervals of mean intensity are comparable in size to the data markers.

**Table S8. Statistical variables**

| variable | type | symbol | distribution | parameters |
|---|---|---|---|---|
| nanofluidic depth | dimension | $d_i$ | uniform | lower and upper bounds: $d_i$ and $d_{i+1}$ |
| root-mean-square surface roughness of replicas | dimension | $R_{q,r}$ | normal | mean: 0 nm, s.d.: 0.74 nm |
| root-mean-square surface roughness of coverslips | dimension | $R_{q,c}$ | normal | mean: 0 nm, s.d.: 0.77 nm |
| calibration error of atomic-force microscope | uncertainty | $\varepsilon_{calibration,i}$ | normal | mean: 0 nm, s.d.: $0.0025 \cdot d_i$ |
| roughness errors of atomic-force microscope | uncertainty | $\varepsilon_{roughness}$ | normal | mean: 0 nm, s.d.: 0.030 nm |
| flatness errors of atomic-force microscope | uncertainty | $\varepsilon_{flatness}$ | normal | mean: 0 nm, s.d.: 0.065 nm |
| nanoparticle radii | dimension | $a_i$ | empirical | $d_i, R_{q,r}, R_{q,c}, \varepsilon_{calibration,i}, \varepsilon_{scan}, \varepsilon_{flatness}$ |
| shape parameter of lognormal distribution of step-edge width | uncertainty | $s_{shape}$ | normal | mean: 0.72, s.d.: 0.035 |
| scale parameter of lognormal distribution of step-edge width | uncertainty | $s_{scale}$ | normal | mean: 432 nm, s.d.: 21 nm |
| step-edge widths | dimension | $\sigma_{se}$ | lognormal | shape: $s_{shape}$, location: 0, scale: $s_{scale}$ |
| step edges | dimension | $\xi_i$ | normal | mean: 0 nm, s.d.: $\sigma_{se}$ |
| brightfield image pixel size (peak wavelength 460 nm) | dimension | $a_{460\,nm}$ | uniform | range.: 99.95 nm to 100.15 nm |
| fluorescence image pixel size (peak wavelength 515 nm) | dimension | $a_{515\,nm}$ | uniform | range: 100.53 nm to 100.74 nm |
| slope of line of best fit to fiducials | dimension | $c_{1,i}$ | student $t$ | degrees of freedom: 3 |
| offset of line of best fit to fiducials | dimension | $o_i$ | student $t$ | degrees of freedom: 3 |
| positions of fiducials | dimension | $r_{f,i}$ | normal | mean: $(x_{f,i}, y_{f,i})$, s.d.: $\sigma_{f,i}$ |
| theoretical localization precision of fiducials | uncertainty | $\sigma_{f,i}$ | normal | mean: $r_{f,i}$, s.d.: $\sigma_{f,i}$ |
| positions of nanoparticles | dimension | $r_j$ | normal | mean: $(x_j, y_j)$, s.d.: $\sigma_j$ |
| theoretical localization precision of nanoparticles | uncertainty | $\sigma_{np,j}$ | normal | mean: $r_j$, s.d.: $\sigma_j$ |
| horizontal offset of size-exclusion regions | dimension | $x_{offset,i}$ | empirical | $R_i, \xi_i, c_{1,i}$ |
| proximity of nanoparticle to centers of size-exclusion regions | dimension | $\mathfrak{D}_{ij}$ | empirical | $c_{1,i}, x_{offset,i}, r_j, o_i$ |
| fluorescence intensity of single nanoparticles | dimension | $I_j$ | poisson | variance: $I_j$ |
| background signal in fluorescence micrographs | dimension | $I_{bkg}$ | poisson | variance: $I_{bkg}$ |
| fluorescence intensity of fluorophore solution in the staircase | dimension | $I_{fs}$ | normal | mean: $\hat{I}_{fs}$, s.d.: $\sigma_{fs,N}$ |
| fluorescence intensity of the interference calibration | dimension | $\hat{I}_{calibration,i}$ | empirical | $I_{fs}, d_i$ |

We abbreviate standard deviation in this table as s.d.
We apply mean values of slopes and offsets of lines of best fit to fiducials to Student $t$-distributions by summation.
We use Student $t$ distributions if the number of degrees of freedom is less than 30.
$r_{f,i} = (x_{f,i}, y_{f,i})$ denotes x and y positions of fiducials corresponding to the $i^{th}$ step.
$r_j = (x_j, y_j)$ denotes the position of the $j^{th}$ nanoparticle.
$\sigma_{f,i}$ denotes Cramér-Rao lower bounds of the localization precision of fiducials.
$\sigma_j$ denotes the Cramér-Rao lower bounds of the localization precision of nanoparticles.[21]

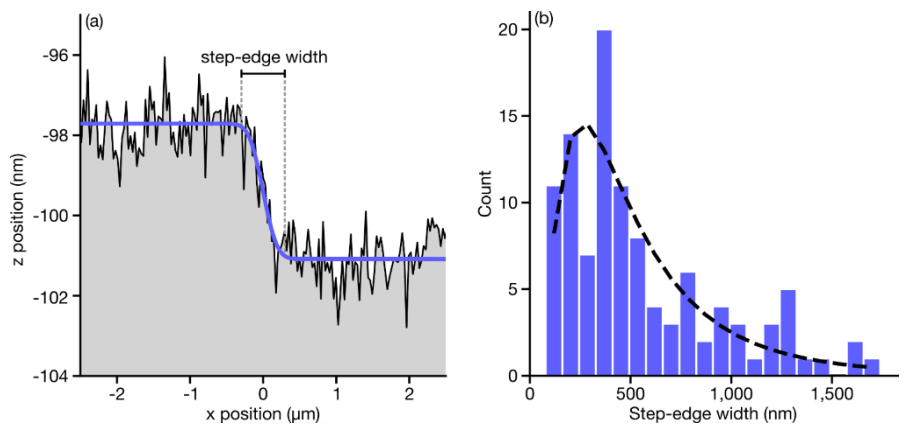

**Figure S17. Step-edge widths.** (a), Atomic-force micrograph section showing a representative step edge of a silicone replica (black). Fitting an error function (blue) to the profile, $z_{i+1} = z_i + \Delta d/2 \cdot \text{erf}\left[(x - x_0)/(\sqrt{2}\sigma_{se,i})\right]$ by damped least-squares with uniform weighting, where $\Delta d$ is the step-depth increment, $x_0$ is the x position of the step, and $\sigma_{se,i}$ is the standard deviation of the Gaussian function corresponding to the error function, determines $4\sigma_{se,i}$ as a 95 % coverage interval for the width of the $i^{th}$ step edge. Analysis of this representative step edge results in a width of 595 nm ± 96 nm. (b), Histogram showing step-edge widths throughout the replica. Fitting a lognormal distribution by maximum likelihood estimation (black dash line) with location 0 to the histogram yields a shape parameter of 0.72 ± 0.07 and a scale parameter of 432 nm ± 42 nm. The location, shape, and scale parameters define the distribution of step-edge width, $\sigma_{se,i}$, for the entire device.

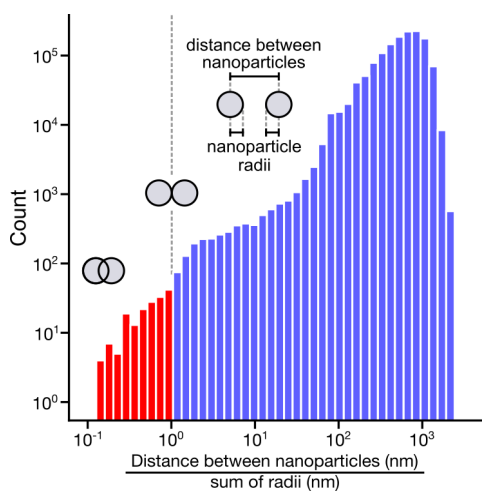

**Figure S18. Steric filter.** Histogram showing the ratio of distances between nanoparticles to the sum of the radii of unique nanoparticle pairs in size-exclusion regions. A steric filter rejects nanoparticle pairs with positions that yield (red bars) distances between nanoparticle that are less than the sum of the nanoparticle radii. We retain nanoparticles with positions that yield ratios of distances between nanoparticles to the sum of the radii that are (blue bars) greater or equal to unity.

**Table S9. Filter summary**

| filter | emitter count | emitters filtered | fraction of emitters filtered | fraction of single nanoparticles filtered |
|---|---|---|---|---|
| none | 10,072 | 0 | 0.000 | – |
| low intensity | 6,996 | 3,076 | 0.305 | – |
| replicate | 4,791 | 2,205 | 0.219 | – |
| point spread function | 3,105 | 1,686 | 0.167 | – |
| steric | 2,761 | 344 | 0.034 | 0.125 |
| size exclusion | 801 | 1,960 | 0.195 | 0.710 |
| reference distribution bounds | 708 | 93 | 0.009 | 0.034 |

Emitter counts are from all four comparable experiments in Figures S19.
We consider single nanoparticles (white region of table) to be those emitters that pass the low intensity, replicate, and point spread function filters.
Emitter count may vary by a few counts with respect to particle numbers in Tables S7, S12, and S13 due to uncertainty propagation by bootstrap resampling.

**Table S10. Analytical yield**

| experiment | nanoparticle count | nanoparticles in size-exclusion regions | spurious yield | analytical yield |
|---|---|---|---|---|
| 1 | 540 | 186 | 65.6 % | 34.4 % |
| 2 | 460 | 126 | 72.6 % | 27.4 % |
| 3 | 581 | 179 | 69.2 % | 30.8 % |
| 4 | 1,180 | 310 | 73.7 % | 26.3 % |
| | | mean | 70.3 % | 29.7 % |
| | | standard deviation | 3.7 % | 3.7 % |

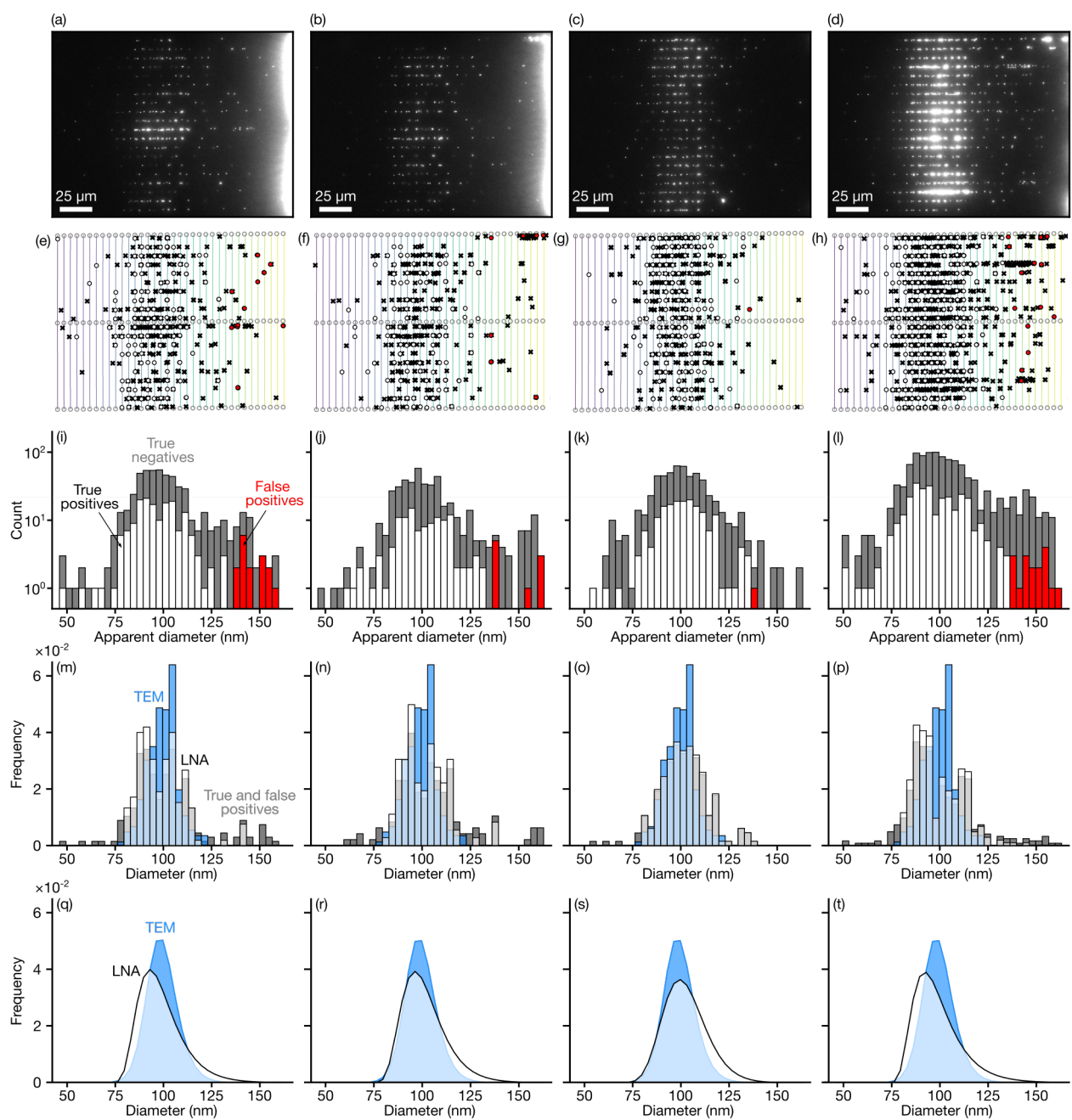

**Figure S19. Comparable experiments.** (a-d) Fluorescence micrographs showing the size separation of nanoparticles in silicone devices. (e-h) Plots showing nanoparticle and fiducial positions and size-exclusion regions. (i-l) Histograms showing diameters apparent from device depths for nanoparticles (gray) outside of size-exclusion regions, (white) inside of size-exclusion regions, and (red) inside of size-exclusion regions but beyond an upper bound. (m-p) Histograms showing diameter distributions for (blue) reference measurements by transmission electron microscopy (TEM), (gray) size-exclusion including both true and false positives, and (white) measurements by our lateral nanoflow assay (LNA), including correction from proior information of the upper and lower bound of the reference distribution. (q-t) Fits of the Johnson $S_U$ distribution to the experimental diameter histograms in (m-p), which we show to guides the eye. Parameters of best fit for the reference histogram and the histograms for each experiment are in Table S11. Columns show the results of experiments that differ by both exposure time to oxygen plasma prior to bonding of silicone replicas to silica coverslips and by the time between device wetting and fluorescence microscopy: (a) exposure to oxygen plasma for 30 s and microscopy approximately 6 h after wetting, (b) exposure to oxygen plasma for 15 s and microscopy approximately 6 h after wetting, (c) exposure to oxygen plasma for 5 s and microscopy 6 h after wetting, and (d) exposure to oxygen plasma for 5 s and microscopy approximately 10 h after wetting.

**Table S11. Fit results**

|  | fit parameters of the Johnson $S_U$ distribution | | | | reduced chi-squared, $\chi_\nu^2$ |
|---|---|---|---|---|---|
|  | shape, $\gamma$ (nm) | shape, $\delta$ (nm) | scale, $\lambda$ (nm) | location, $\xi$ (nm) |  |
| reference measurements | -0.906 | 2.69 | 92.0 | 19.8 | 7.9 |
| experiment 1 | -7.92 | 2.55 | 69.4 | 2.44 | 2.6 |
| experiment 2 | -8.94 | 2.91 | 68.5 | 2.87 | 2.6 |
| experiment 3 | -10.2 | 5.00 | 47.7 | 14.2 | 1.3 |
| experiment 4 | -8.09 | 2.25 | 71.5 | 1.37 | 6.6 |
| all experiments | -0.482 | 1.26 | 92.4 | 14.2 | 5.2 |

We plot fits of the Johnson $S_U$ distribution to the data from all experiments in Figure 4.

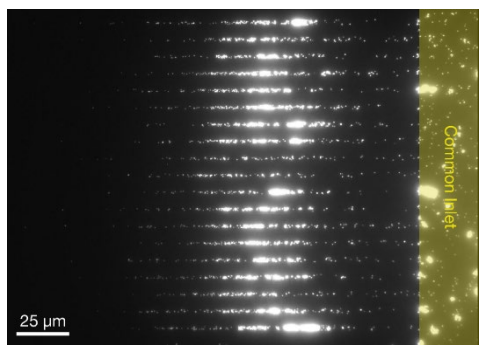

**Video S1. Surface stability.** A time series of fluorescence micrographs showing the Brownian motion in real time of some nanoparticles in deeper steps of staircase structures. This motion occurs 100 h after wetting the disposable device, which is near the end of the useful duration for capillarity to drive hydrodynamic transport. The presence of Brownian motion indicates the ongoing mitigation of attractive interactions between nanoparticles and confining surfaces, demonstrating the stability of the system at that time scale. These data are for the longest exposure to oxygen plasma of 30 s. The yellow region indicates the common inlet of the device.

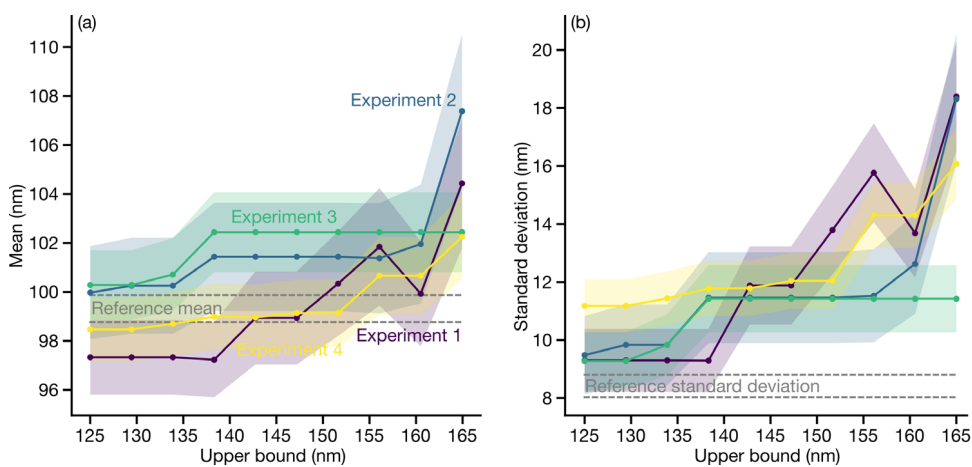

**Figure S20. Sensitivity analysis.** Plots showing (a) mean diameter and (b) standard deviation of measurements by the lateral nanoflow assay with correction as a function of upper bound for each experiment. Solid circles represent mean values and shaded regions indicate 95 % coverage intervals. The dash gray lines in either plot indicate the 95 % coverage interval of the (a) mean and (b) standard deviation from the reference diameter distribution.

Table S12. Measurement summary

| experiment | calibration | relevant bounds for correction | | number of particles | moments of the diameter histogram | | | | | moments of the marginal intensity histogram | |
|---|---|---|---|---|---|---|---|---|---|---|---|
| | | lower (nm) | upper (nm) | | mean (nm) | error in mean (nm) | (%) | standard deviation (nm) | error in standard deviation (nm) | (%) | shape | scale |
| **reference** | – | **0** | **∞** | 668 | **99.3** | – | – | **8.4** | – | – | – | – |
| all | none | 78.1 | 134.3 | 668 | 98.6 ± 1.5 | -0.8 | -0.8 | 10.0 ± 1.2 | 1.6 | 18.9 | 0.79 ± 0.15 | 0.68 ± 0.11 |
| all | flatfield | 78.1 | 134.3 | 670 | 98.5 ± 1.4 | -0.8 | -0.8 | 9.8 ± 1.1 | 1.4 | 16.4 | 0.79 ± 0.14 | 0.69 ± 0.06 |
| all | PSF | 78.1 | 134.3 | 527 | 98.7 ± 2.0 | -0.6 | -0.6 | 10.4 ± 1.4 | 2.0 | 23.8 | 0.63 ± 0.06 | 0.84 ± 0.08 |
| all | position | 78.1 | 134.3 | 794 | 98.9 ± 0.9 | -0.4 | -0.4 | 10.3 ± 0.5 | 1.8 | 21.9 | 0.86 ± 0.14 | 0.62 ± 0.08 |
| all | interference | 78.1 | 134.3 | 667 | 98.6 ± 1.5 | -0.7 | -0.7 | 10.0 ± 1.2 | 1.6 | 18.8 | 0.82 ± 0.16 | 0.68 ± 0.11 |
| 1 | all | 78.1 | 134.3 | 135 | 97.0 ± 1.7 | -2.3 | -2.3 | 9.9 ± 1.2 | 1.5 | 18.0 | 0.67 ± 0.09 | 0.75 ± 0.13 |
| 2 | all | 78.1 | 134.3 | 86 | 99.6 ± 2.2 | 0.3 | 0.3 | 10.1 ± 1.3 | 1.7 | 20.3 | 0.62 ± 0.10 | 0.79 ± 0.14 |
| **3** | **all** | **78.1** | **134.3** | 153 | **100.5 ± 1.7** | **1.2** | **1.2** | **10.4 ± 1.2** | 2.0 | 23.8 | **0.72 ± 0.08** | **0.90 ± 0.10** |
| 4 | all | 78.1 | 134.3 | 246 | 98.8 ± 1.5 | -0.5 | -0.5 | 11.9 ± 0.9 | 3.5 | 41.9 | 0.71 ± 0.05 | 0.88 ± 0.10 |
| all | all | 0.0 | ∞ | 799 | 102.1 ± 1.3 | 2.8 | 2.8 | 17.7 ± 2.5 | 9.3 | 110.3 | 0.73 ± 0.05 | 0.82 ± 0.05 |
| all | all | 78.1 | ∞ | 752 | 104.0 ± 1.5 | 4.7 | 4.7 | 16.3 ± 2.5 | 7.9 | 94.2 | 0.69 ± 0.04 | 0.86 ± 0.04 |
| **all** | **all** | **0.0** | **134.3** | 669 | **97.0 ± 1.6** | **2.3** | **2.3** | **12.6 ± 0.8** | 4.2 | 50.0 | **0.73 ± 0.05** | **0.78 ± 0.07** |
| **all** | **all** | **78.1** | **134.3** | 620 | **98.9 ± 1.3** | **-0.4** | **-0.4** | **10.6 ± 0.8** | 2.2 | 25.9 | **0.68 ± 0.04** | **0.83 ± 0.06** |

We abbreviate point-spread function in this table as PSF.
We calculate errors and percent errors by comparison to the reference diameter histogram from transmission electron microscopy.
We calculate shape the standard deviation of the natural log of nanoparticle intensities. We calculate scale as the median of nanoparticle intensities.
Relevant lower and upper bounds of 0 and ∞ correspond to measurements with no corrections.
Numbers of particles may vary by a few particles with respect to Tables S7, S9, and S13 due to uncertainty propagation by bootstrap resampling.
Bold rows indicate the subset of data that we show in the main text.

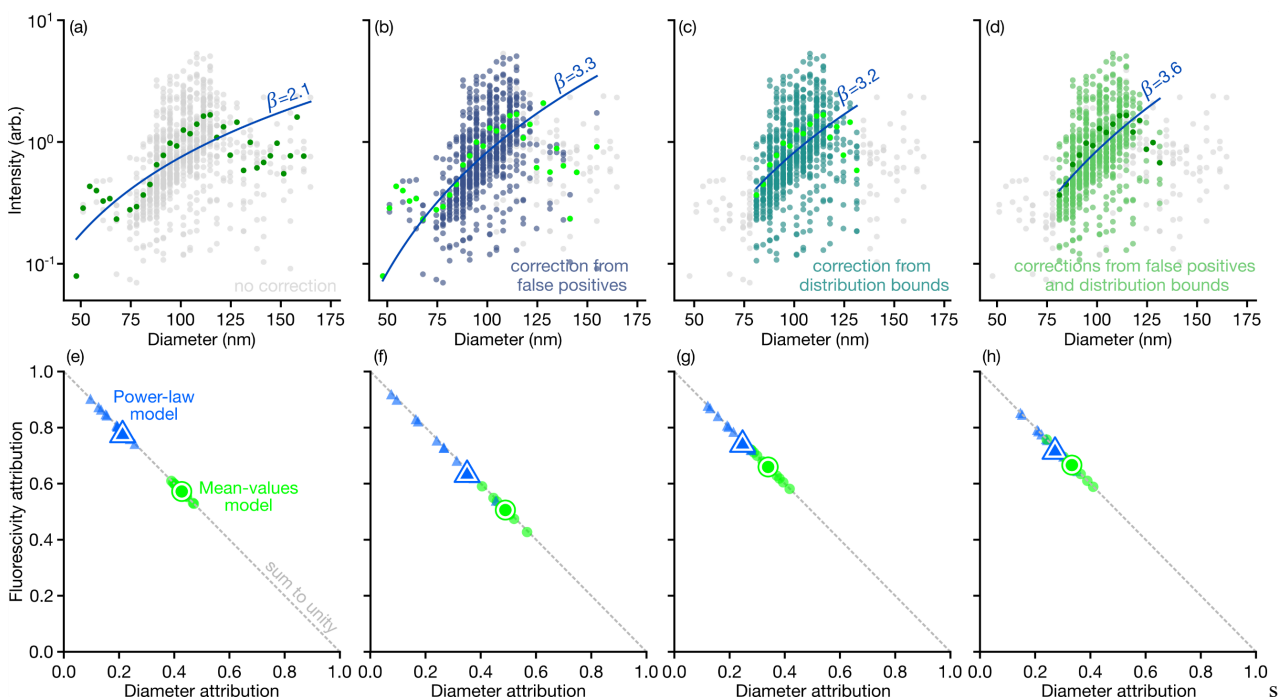

**Figure S21. Bayesian statistical analysis.** (a-d) Plots showing fluorescence intensity as a function of steric diameter for various aspects of our sizing correction, which can make use of different amounts of prior information, including (a) no correction, (b) correction from false positives above the upper bound of the reference diameter distribution, (c) correction from both the lower and upper bound of the reference diameter distribution, and (d) correction from both false positives above the upper bound and the lower and upper bounds of the reference diameter distribution. Dark green circles are mean values of fluorescence intensity for each diameter and solid blue lines are power-law trends. Grey circles in represent data without correction. Further details of the power-law parameters are in Table S14. (e-f) Plots showing the fractions of intensity heterogeneity that are attributable to diameter and fluorescivity variability for two Bayesian models for each correction. Green circles are the mean-values model. Blue triangles are the power-law model. Data markers are the means of posterior distributions. Solid lines are the major axes, or approximately 95 %, of these distributions of representative points. The minor axes are comparable to the line widths. The dash gray line indicates sum to unity, with a slight distance to the data markers being attributable to measurement uncertainty. Further details of the attribution of fluorescivity variability are in Table S15.

**Table S13. Posterior predictive checks**

| | | posterior predictive check | |
|---|---|---|---|
| correction | number of particles | mean-values model | power-law model |
| none | 802 | 0.055 | 0.056 |
| false positives | 681 | 0.057 | 0.051 |
| distribution bounds | 703 | 0.055 | 0.058 |
| both above | 622 | 0.061 | 0.059 |

The target in all cases is 0.05.
Numbers of particles may vary by a few particles with respect to Tables S7, S9, and S12 due to uncertainty propagation by bootstrap resampling.

**Table S14. Power-law parameters**

| | parameters of the power-law model | |
|---|---|---|
| correction | scale, $\alpha$ | exponent, $\beta$ |
| none | -9.9 ± 2.1 | 2.1 ± 0.5 |
| false positives | -15.3 ± 2.2 | 3.3 ± 0.5 |
| distribution bounds | -15.1 ± 2.4 | 3.2 ± 0.5 |
| both above | -16.8 ± 2.5 | 3.6 ± 0.5 |

We report the scale parameter in log space.
Raising the Euler number, $e$, to the value of the scale parameter transforms the scale parameter into linear space, $e^{\alpha}$.
We compute reduced chi-square statistics for the power-law model relative to mean values and variances from each diameter bin in Figure S21.

**Table S15. Sources of fluorescivity variability**

| | bounds (nm) | | diameter attribution | | | | fluorescivity attribution | | | | uncertainty attribution | | | |
|---|---|---|---|---|---|---|---|---|---|---|---|---|---|---|
| | | | power law | | mean values | | power law | | mean values | | power law | | mean values | |
| correction | lower | upper | mean | s.d. | mean | s.d. | mean | s.d. | mean | s.d. | mean | s.d. | mean | s.d. |
| none | 0 | ∞ | 0.213 | 0.046 | 0.427 | 0.050 | 0.787 | 0.046 | 0.572 | 0.050 | $3.5\times10^{-4}$ | $1.2\times10^{-4}$ | $3.2\times10^{-4}$ | $9.7\times10^{-5}$ |
| false positives | 0 | 135 | 0.355 | 0.070 | 0.494 | 0.044 | 0.645 | 0.070 | 0.506 | 0.044 | $3.5\times10^{-4}$ | $1.2\times10^{-4}$ | $3.0\times10^{-4}$ | $8.3\times10^{-5}$ |
| distribution bounds | 78 | 135 | 0.247 | 0.054 | 0.340 | 0.043 | 0.752 | 0.054 | 0.660 | 0.043 | $3.7\times10^{-4}$ | $1.2\times10^{-4}$ | $3.6\times10^{-4}$ | $7.8\times10^{-5}$ |
| both above | 78 | 135 | 0.272 | 0.049 | 0.333 | 0.043 | 0.728 | 0.049 | 0.667 | 0.043 | $3.8\times10^{-4}$ | $1.2\times10^{-4}$ | $3.7\times10^{-4}$ | $9.2\times10^{-5}$ |

We abbreviate standard deviation in this table as s.d.